\documentclass[aps,pra,superscriptaddress,twocolumn,10pt]{revtex4-1}
\usepackage{times,xcolor,amsmath,amsfonts,amsthm,graphicx,qcircuit}
\usepackage[colorlinks = true,breaklinks = true]{hyperref}

\newcommand{\RR}{\mathbb{R}}
\newcommand{\CC}{\mathbb{C}}
\newcommand{\ra}{\rangle}
\newcommand{\la}{\langle}

\newcommand{\revision}[1]{{\color{black}#1}}

\begin{document}


\author{Bela Bauer}
\affiliation{Microsoft Quantum, Station Q, University of California, Santa Barbara, CA 93106, USA}
\email{Bela.Bauer@microsoft.com}

\author{Sergey Bravyi}
\affiliation{IBM T. J. Watson Research Center, Yorktown Heights, NY 10598, USA}
\email{sbravyi@us.ibm.com}

\author{Mario Motta}
\affiliation{IBM Almaden Research Center, San Jose, CA 95120, USA}
\email{mario.motta@ibm.com}

\author{Garnet Kin-Lic Chan}
\affiliation{Division of Chemistry and Chemical Engineering, California Institute of Technology, Pasadena, CA 91125, USA}
\email{garnetc@caltech.edu}

\title{Quantum algorithms for quantum chemistry and quantum materials science}

\begin{abstract}
  As we begin to reach the limits of classical computing, quantum computing has emerged as a technology
  that has captured the imagination of the scientific world. While for many years, the ability to execute quantum
  algorithms was only a theoretical possibility, recent advances in hardware mean that quantum computing devices
  now exist that can carry out quantum computation on a limited scale. Thus it is now a real possibility, and of
  central importance at this time, to assess the potential impact of quantum computers on real problems of interest.
  One of the earliest and most compelling applications for quantum computers is Feynman’s idea of simulating
  quantum systems with many degrees of freedom. Such systems are found across chemistry, physics, and materials science. The particular way in which quantum computing extends classical computing means that one
  cannot expect arbitrary simulations to be sped up by a quantum computer, thus one must carefully identify areas
  where quantum advantage may be achieved. In this review, we briefly describe central problems in chemistry
  and materials science, in areas of electronic structure, quantum statistical mechanics, and quantum dynamics,
  that are of potential interest for solution on a quantum computer. We then take a detailed snapshot of current
  progress in quantum algorithms for ground-state, dynamics, and thermal state simulation, and analyze their
  strengths and weaknesses for future developments.
\end{abstract}

\maketitle

\section{What is a quantum computer? \\ How is it relevant to quantum simulation?}
\label{sec:intro}

\revision{A quantum computer is a device which expands the computational capabilities of a classical computer
  via the processing of quantum information}~\cite{nielsen2000quantum,mermin2007quantum,benenti2004principles,kitaev2002classical,yanofsky2008quantum,marinescu2005approaching}. 
The basic unit of quantum information, called a \textit{qubit}, is synonymous with a two-level quantum system.
Denoting the two basis states of a qubit as $|0\rangle,|1\rangle$,
the general single-qubit state 
may be a superposition \revision{$|\psi \rangle = \sum_{z=0,1} c_z | z \rangle$ where $c_0$, $c_1$ are complex numbers satisfying $\sum_z |c_z|^2=1$ (see
Figure \ref{fig:bloch_sphere} for a common visualization).
For $n$ qubits, there are $2^n$ basis states, which can be enumerated as bitstrings $|z_1 \dots z_n \rangle$, $z_1 \dots z_n = 0,1$. To specify a general 
quantum state of $n$ qubits, one must specify a complex coefficient (amplitude) for each basis state, i.e. $| \psi \rangle = \sum_{z_1 \dots z_n = 0,1} c_{z_1 \dots z_n} |z_1 \dots z_n \rangle$. The exponential number of amplitudes
needed to specify the general state of $n$ qubits stands in contrast to the linear amount of information needed to encode a single bitstring $|z_1 \dots z_n \rangle$ (each of which is a state of $n$ classical bits)}.

\revision{
\begin{figure}[b]
\centering
\includegraphics[width=0.55\columnwidth]{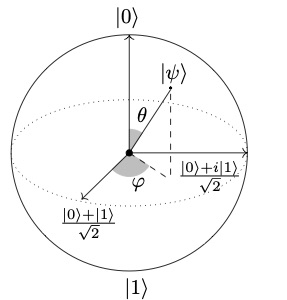}
\caption{
A common way
  to visualize the state of a single-qubit is to parametrize it as $|\psi\rangle = \cos(\theta/2) | 0 \rangle + e^{i\phi} \sin(\theta/2) | 1 \rangle$
  where the angles $\theta, \phi$ 
  map the state onto a point on the surface of a sphere, known as the Bloch sphere.
    The north and south poles, $|0\rangle$ and $|1\rangle$, represent the ``classical states'' (or computational basis states) and denote the bits $0, 1$ used in a classical computer.
}
\label{fig:bloch_sphere} 
\end{figure}
}

\revision{
Measuring the qubits in this basis, often referred to as the computational basis, collapses the state onto a single (random) bitstring $|z_1, \ldots z_\rangle$ with the probability $|c_{z_1 \ldots z_n}|^2$. Hence, the description of quantum mechanical processes is always probabilistic. However, quantum mechanics is not completely captured by classical probabilities. In particular, instead of measuring each qubit in the computational basis, we can measure in a different basis; in the visualization of Fig.~\ref{fig:bloch_sphere}, this corresponds to measuring not along the $\hat{z}$ axis, but for example the $\hat{x}$ axis. Such a measurement will again yield a single bit string in the new basis, $|x_1 \ldots x_n\rangle$ with probability $|c_{x_1 \ldots x_n}|^2$. However, this new probability is a function of the full complex amplitudes of the state and cannot be obtained in terms of a single classical probability distribution.
This gives rise to the possibility for storing more general kinds of correlations in the quantum state~\cite{bell1964einstein,clauser1969proposed}, which is
the heart of the quantum phenomenon of entanglement, the name given to correlations in a system that
cannot be mimicked by a classical distribution of states.
The possibility of creating entangled states in a space of exponentially large dimension
and manipulating these states by exploiting their constructive or destructive interference 
is the ultimate source of the computational power of a quantum computer.
}
However, despite the exponential separation between quantum and classical information, this does
not simply mean that quantum computers can compute answers to problems ``in parallel'' with
an exponential speedup, for example, by storing multiple different solutions in the many amplitudes.
This is because reading out from a quantum computer destroys the state, and thus to harness the power
of quantum information, a quantum algorithm must coordinate interference between the amplitudes such that
useful information can be read out with high confidence without \revision{exponentially many measurements}.

The interest in quantum computing for quantum simulations of molecules and materials
stems from the fact that in many cases,
the chemistry and physics of molecules and materials is best described using quantum mechanics.
Thus, the state of a many-particle molecule also encodes quantum information, and as the number
of atoms increases, similarly can require an exponentially large number of classical bits to describe.
This means that in the worst case, quantum simulation is exponentially hard on classical computers.
This is the motivation for Feynman's famous observation that ``Nature isn't classical, dammit, and if you 
want to make a simulation of nature, you'd better make it quantum mechanical''~\cite{feynman1982simulating}.

A moment's reflection, however, suggests that the potential quantum advantage for a quantum computer 
in quantum simulation is nonetheless subtle. For example, if it were indeed impossible to say anything
about how atoms, molecules, or materials behave, without using a quantum computer, there would be no 
disciplines of chemistry, condensed matter physics, or materials science! 
Decades of electronic structure and quantum chemistry simulations suggest that reasonable, and in some 
cases very high accuracy, solutions of quantum mechanics can be obtained by classical algorithms in 
practice. Quantum advantage in quantum simulation is thus problem-specific, and must be tied both to the 
types of questions that are studied, as well as the accuracy required.

We can look to theoretical quantum computer science to better understand the power of quantum computers 
in quantum simulation.
The natural problem to solve on a quantum computer is the time evolution of a quantum system given some 
initial state,
\begin{align}
\textit{Quantum dynamics:} \, & \, i \partial_t |\Psi(t)\rangle = \hat{H} |\Psi(t)\rangle.
\end{align}
This problem is representative for the complexity class $\mathsf{BQP}$ \revision{(bounded-error quantum polynomial)} i.e. it is of polynomial cost on a quantum 
computer and believed to offer a clear separation from the classical case (although an exponential quantum speedup has 
been rigorously proven only in the query complexity setting~\cite{mosca1998hidden}).
However, it is necessary to prepare the initial  state, which may be difficult by itself. In particular, preparing a 
low-energy state may be challenging, which naturally leads to considering other important problems,
\begin{align}
\textit{Ground state:} \, & \, E_0 = \min_{|\Psi\rangle} \langle \Psi |\hat{H}|\Psi\rangle \\
 \textit{Thermal averages:} \, & \, \langle \hat{A} \rangle = \frac{\mathrm{Tr}\big[ \hat{A} e^{-\beta \hat{H}} \big]}{\mathrm{Tr}\big[ e^{-\beta \hat{H}} \big]}
\end{align}
Ground state determination lies in complexity class $\mathsf{QMA}$ \revision{(quantum Merlin-Arthur)} \cite{kitaev2002classical}, 
a class of problems not known to be efficiently 
solvable (in general) on a quantum computer~\footnote{\revision{$\mathsf{QMA}$ is known to contain the computational complexity classes $\mathsf{NP}$ (nondeterministic polynomial time, often
    considered to be a class of problems not known to be efficiently solvable on classical computers) and $\mathsf{BQP}$, but it is unknown whether these
inclusions are strict.}}.
This also means that thermal averages cannot in general be 
computed efficiently on a quantum computer, since in the limit of zero temperature, this problem reduces to 
ground-state determination.
Although it is the case that there are many physical ground state and thermal problems that are not hard to 
solve in practice (as demonstrated by efficient classical simulation of many problems) and similarly many initial states of interest in quantum
dynamics that are easy to prepare, the above tells us that in a rigorous sense, we do not have a
theoretical guarantee that
a quantum computational advantage can be achieved for the central problems in quantum
simulation.

Given the limits to the guidance that can be provided by rigorous computational complexity results, it is clear that 
to understand quantum advantage in chemistry, condensed matter physics, and quantum materials science, 
we must be guided by actual empirical data in the form of numerical and theoretical experiments
with quantum algorithms and quantum devices on simulation problems of interest. This requires making progress 
on both theoretical and practical questions of quantum simulations, ranging from the basic algorithms and choices 
of encoding and representation to issues of circuit compilation, readout, and mapping to specialized hardware. 
A central purpose of this review is to provide a perspective on what the relevant 
chemical and materials problems are today; to give a snapshot of the limitations of classical methods for these 
problems; and in these contexts to understand the strengths, weaknesses, and bottlenecks of existing ideas for 
quantum algorithms, and where they need to be improved, both in terms of fundamental theoretical aspects as 
well as practical methods of implementation.

\subsection{Current quantum architectures}
\label{sec:quantum-architectures}

\begin{figure}[ht]
\centering
\includegraphics[width=0.7\columnwidth]{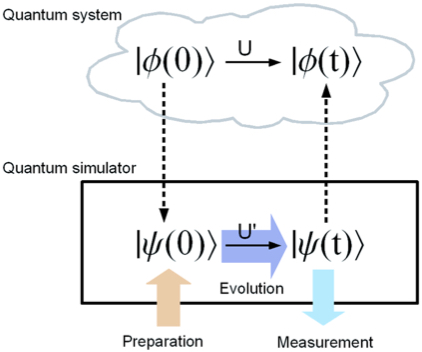}
\caption{Schematic of a quantum simulation of quantum dynamics. A quantum simulator (bottom) is prepared in an initial state $|\psi(0)\rangle$,
that is a representation of the initial state $|\phi(0) \rangle$ of the actual system of interest (top). The simulator is manipulated by a 
unitary transformation $\hat{U}^\prime$ that is an encoding of the real-time evolution of the system of interest, and the final simulator state 
$|\psi(t)\rangle$ is measured, yielding information about the dynamics of the original system. From Ref.~\cite{georgescu2013quantum}.}
\label{fig:simulator} 
\end{figure}

The idea of using a quantum mechanical device to perform a computation was first considered in earnest by Richard P. Feynman in 
a famous lecture in 1982~\cite{feynman1982simulating}.
Feynman's suggestion was to build a lattice of spins with tunable interactions. 
He conjectured that by appropriately tuning these interactions, such a system could be made to imitate the behavior of any other (bosonic) 
quantum system with the same dimensionality, and thus could serve as a way to compute the properties of some other system that one 
would like to study. This idea, which is often \revision{referred to as \textit{analog quantum computation} \cite{solano1,solano2,solano3}}, is still very much alive today and embodied 
in the field of cold atomic gases and related techniques, which has made great progress in simulating complicated physics of strongly 
correlated systems in a controlled environment. A general schematic of the idea is shown in \revision{Figure~}\ref{fig:simulator}. 

However, such systems come with some limitations: \revision{in particular, a given setup for an analog quantum computer can practically
  only simulate a certain subset of physical systems. For example, it would be challenging to tune the physical interactions
  between cold atoms arranged in a one-dimensional lattice to simulate local interactions on a higher-dimensional graph. Furthermore, just as in the
  classical case, analog quantum computers are ultimately limited in accuracy 
because over time, errors in the experiment accumulate without a systematic means of correction.}
In this review, we therefore focus on 
a second,  more general, approach to quantum simulations offered by 
\textit{digital quantum computation}. Here, very much in analogy to the 
classical computers we are used to, one considers a system of quantum registers -- qubits -- that are controlled through some set of 
instructions - the quantum gates.
\revision{Quantum gates are a convenient way to describe and understand the manipulation of quantum states.
Usually the gate operations are defined over only a few qubits at a time,
thus the action of each gate can be viewed as a unitary time evolution 
of the $n$-qubit system under a
suitable Hamiltonian acting non-trivially over only on a few qubits (usually one or two).}
Any quantum computation can be expressed by a sequence of such elementary gates,
called a \textit{quantum circuit}, applied 
to an initial basis state, e.g. $|00 \ldots 0\rangle$, and followed by the measurement of some set of the qubits.
\revision{Some examples of elementary gates used throughout this review are shown in Table \ref{tab:gates}.}
One can show that with a finite (and indeed relatively small) set of such gate operations, 
one can in principle generate arbitrary quantum evolution for a finite time to arbitrary precision!
In other words, every problem in the complexity 
class $\mathsf{BQP}$ can be mapped into a quantum program, i.e. the quantum circuit composed of such elementary gates.

\revision{
  \begin{table*}[t]
\begin{tabular}{|ccc|ccc|}
\hline \hline
name & symbol & matrix representation & name & symbol & matrix representation \\
\hline
Pauli X & \Qcircuit @C=1em @R=.7em {& \gate{X} & \qw } & $\left( \begin{array}{cc} 0 & 1 \\ 1 & 0 \end{array} \right)$ &
X rotation & \Qcircuit @C=1em @R=.7em {& \gate{R_{x}(\theta)} & \qw } & $\left( \begin{array}{rr} \cos\left(\frac{\theta}{2}\right) & -i \sin\left(\frac{\theta}{2}\right) \\ \\ -i \sin\left(\frac{\theta}{2}\right) & \cos\left(\frac{\theta}{2}\right) \end{array} \right)$ \\
&&&&&\\
Pauli Y & \Qcircuit @C=1em @R=.7em {& \gate{Y} & \qw } & $\left( \begin{array}{rr} 0 & -i \\ i & 0 \end{array} \right)$ &
Y rotation & \Qcircuit @C=1em @R=.7em {& \gate{R_{y}(\theta)} & \qw } & $\left( \begin{array}{rr} \cos\left(\frac{\theta}{2}\right) & -\sin\left(\frac{\theta}{2}\right) \\ \\ \sin\left(\frac{\theta}{2}\right) & \cos\left(\frac{\theta}{2}\right) \end{array} \right)$ \\
&&&&&\\
Pauli Z & \Qcircuit @C=1em @R=.7em {& \gate{Z} & \qw } & $\left( \begin{array}{rr} 1 & 0 \\ 0 & -1 \end{array} \right)$ & Z rotation & \Qcircuit @C=1em @R=.7em {& \gate{R_{z}(\theta)} & \qw } & $\left( \begin{array}{rr} e^{ - i \frac{\theta}{2} } & 0 \\ 0 & e^{ i \frac{\theta}{2} } \end{array} \right)$ \\
&&&&&\\
\hline
&&&&&\\
\,\, Hadamard & \Qcircuit @C=1em @R=.7em {& \gate{H} & \qw } & $\frac{1}{\sqrt{2}} \, \left( \begin{array}{rr} 1 & 1 \\ 1 & -1 \end{array} \right)$ &
$\mathsf{cNOT}$ & \Qcircuit @C=1em @R=.7em {& \ctrl{1} & \qw \\ & \targ & \qw } & $\left( \begin{array}{rrrr} 1 & 0 & 0 & 0 \\ 0 & 1 & 0 & 0 \\ 0 & 0 & 0 & 1 \\ 0 & 0 & 1 & 0 \\ \end{array} \right)$ \\
&&&&&\\
S gate & \Qcircuit @C=1em @R=.7em {& \gate{S} & \qw } & $\left( \begin{array}{rr} 1 & 0 \\ 0 & i \end{array} \right)$ &
$\mathsf{cU}$ & \Qcircuit @C=1em @R=.7em {& \ctrl{1} & \qw \\ & \gate{U} & \qw } & $\left( \begin{array}{rrrr} 1 & 0 & 0 & 0 \\ 0 & 1 & 0 & 0 \\ 0 & 0 & u_{00} & u_{01} \\ 0 & 0 & u_{10} & u_{11} \\ \end{array} \right)$ \\
&&&&&\\
\hline
&&&&&\\
T & \Qcircuit @C=1em @R=.7em {& \gate{T} & \qw } & $\left( \begin{array}{rr} 1 & 0 \\ 0 & e^{i \frac{\pi}{4}} \end{array} \right)$ &
\,\, Z measurement & \Qcircuit @C=1em @R=.7em {& \meter \\ } & $p(k) = | \langle k | \Psi \rangle |^2$ \\
&&&&&\\
\hline
\hline
\end{tabular}
\caption{Examples of quantum gate operations and circuit elements. Top, left: single-qubit Pauli operators. Top, right: single-qubit rotations (for example, $R_x(\theta) = e^{ - i \frac{\theta}{2} X}$. Middle, left: relevant single-qubit operations in the Clifford group. Middle, right: two-qubit $\mathsf{CNOT}$ (controlled-X) and $\mathsf{cU}$ (controlled-U) gates. Bottom, left: T gate. Bottom, right: measurement of a single qubit in the computational basis.}
\label{tab:gates}
\end{table*}
}

A third model of quantum computation is \textit{adiabatic quantum computation}. Here, the computation is encoded into a time-dependent 
Hamiltonian, and the system is evolved slowly to track the instantaneous ground state of this Hamiltonian. It can be shown that this model 
is equivalent to  circuit-based digital quantum computation~\cite{aharonov2008adiabatic}, but it is usually considered to be less practical. 
However, a restricted version of it, adiabatic quantum optimization~\cite{farhi2001quantum}, has \revision{attracted significant attention}.
 Here, a  classical optimization problem is encoded into a quantum Hamiltonian, to which one adds some non-commuting terms to endow the system 
with non-trivial quantum dynamics. If one then slowly turns off the quantum terms, the optimal solution to the classical problem should be 
obtained. In practice, however, one may have to go impractically slowly for this to be true; the detailed analysis of this approach is a 
complex problem that is not covered in this review.

\subsection{Building a circuit-based digital quantum computer}
\label{sec:building}

Returning to circuit-based quantum simulation, what is the status of the field today? The natural enemy of quantum computation is decoherence of the qubit, i.e. the tendency of the stored quantum state to decay into a classical state. After decades of research, a number of qubit technologies \cite{johnson2019multiferroic,kintzel2018molecular}, for example superconducting qubits~\cite{Wendin_2017} and ion traps~\cite{bruzewicz2019trapped}, have reached the point where small devices of a few dozen qubits can be sufficiently isolated from decoherence to execute non-trivial quantum algorithms of a few tens to hundreds of gates. This has been termed the \textit{noisy intermediate-scale quantum} (NISQ) era~\cite{Preskill2018NISQ}. With these devices, the first demonstrations of \textit{quantum supremacy}
appear to be possible. For this, an artificial but well-defined problem that is intractable classically  is solved on a quantum computer~\cite{arute2019quantum}.
\revision{(Here, both ``solve'' and ``classically intractable'' are meant in the practical sense of not being solvable on current classical hardware
  with current algorithms, but where a result can be obtained using current quantum hardware.)}
However, so far the proposed problems do not generally have practical relevance, and a more important and challenging question is when a scientifically or commercially relevant problem can be solved on a quantum computer.

\begin{figure}
    \centering
    \includegraphics[width=0.85\columnwidth]{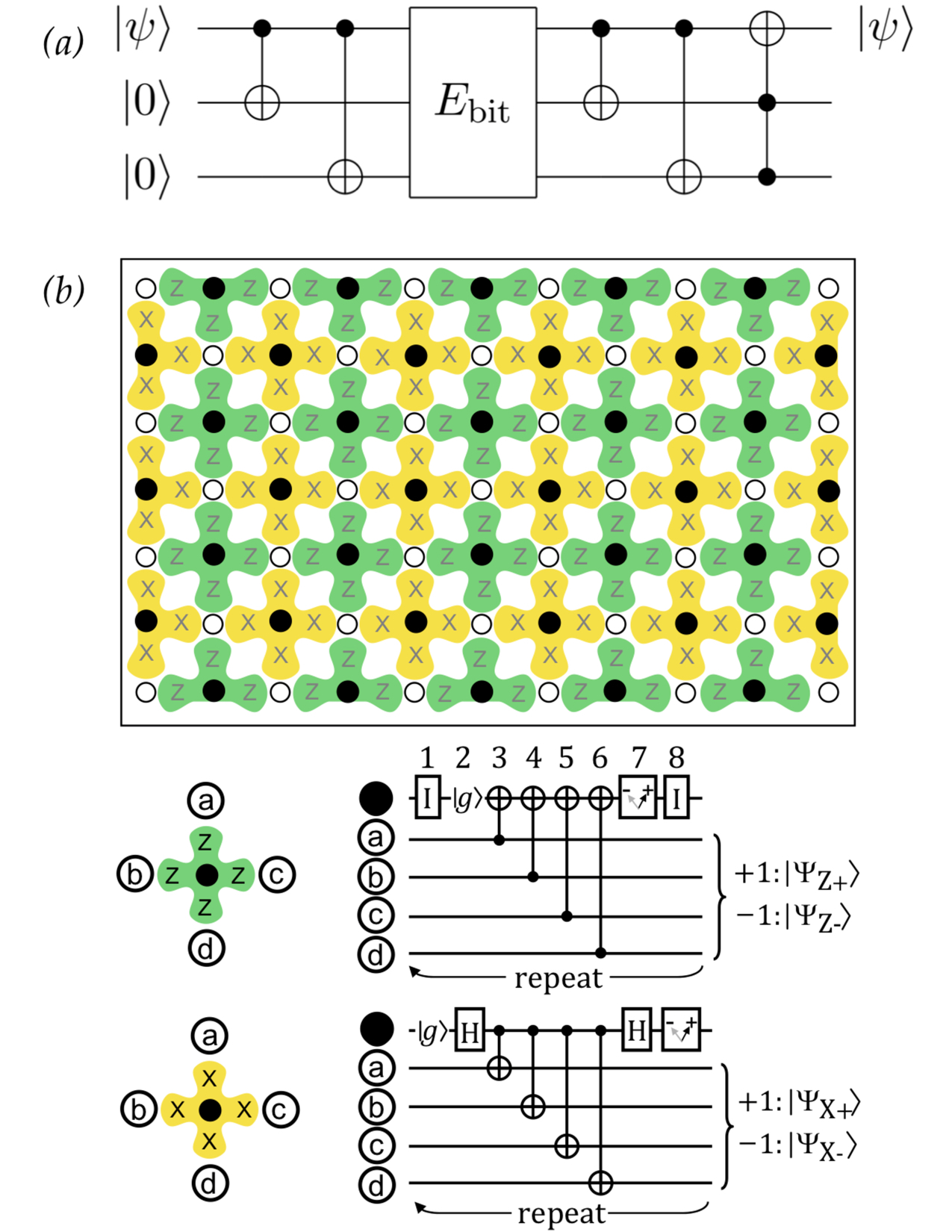}
    \caption{Examples of error correction schemes. (a) Quantum circuit implementing the bit-flip code. The state of a single logical qubit, $| \psi \rangle = \alpha| 0 \rangle + \beta | 1 \rangle$ is encoded in the state of three physical qubits as $\alpha | 000 \rangle + \beta | 111 \rangle$. The system is then sent through a channel $E_{bit} = \mathcal{E}^{\otimes 3}$, where $\mathcal{E}$ flips a qubit with probability $p$ (X error). Measuring the $Z$ operator on the last two qubits permits to determine whether a qubit has been flipped by the channel, and to correct such an error without corrupting the transmitted state.
    (b) Schematic of the surface code for quantum error correction and quantum fault-tolerance. Top: a two-dimensional array implementation of the surface code, data qubits are open circles and measurement qubits are filled circles, with measure-Z qubits colored green (dark) and measure-X qubits colored \revision{yellow} (light). Middle: sequence of operations (left), and quantum circuit (right) for one surface code cycle for a measure-Z qubit, to detect sign flip errors. Bottom: similar, but for measure-X qubits. Adapted from Ref.~\cite{fowler2012towards}.}
    \label{fig:qec}
\end{figure}

\revision{Adressing problems in quantum simulation will almost certainly require larger circuits than current quantum supremacy proposals. To reach these,} it will almost certainly be necessary to correct errors that occur on qubits during the computation. While the no-cloning theorem prevents error correction by simple redundancy, it can be shown that \textit{quantum error correction} is possible nonetheless by encoding a single qubit into an entangled state of many qubits~\cite{terhal2015qec}, to use suitable measurement to detect errors occurring in the system and to apply suitable unitary transformations correcting such errors without disturbing the information encoded in the system (see \revision{Figure~}\ref{fig:qec}). This leads to the very important distinction between \textit{physical} and \textit{logical} qubits. The latter are error-corrected and encoded in the state of many physical qubits. Quantum algorithms are performed on the logical qubits, and the error-correction scheme translates the operations on logical qubits into physical operations.
This incurs significant overhead: depending on the error rate of the underlying physical qubits and the target error rate of the logical qubits, one may need hundreds or even thousands of physical qubits to realize a single logical qubit. Therefore, when evaluating the capabilities of some qubit platform with respect to an algorithm, one must be careful to include the cost of encoding logical qubits into physical qubits to achieve the required error rates.

\section{Simulation challenges in molecular and materials science}
\label{sec:simulations}

\revision{We now turn to survey a representative, but certainly non-exhaustive, set of scientific problems that could be interesting   for quantum simulation. We
  group these roughly into a few areas: \textit{quantum chemistry}, the problem of determining the low-lying states of electrons in molecules; \textit{quantum molecular spectroscopy}, concerned with stationary states of the nuclei vibrating and rotating in molecules; \textit{chemical quantum dynamics},
  that studies the non-equilibrium electronic and nuclear motion of molecules associated with reactions and external fields; \textit{correlated electronic structure in materials},
a close cousin of quantum chemistry in the materials domain, but with several important differences in practice and in emphasis; and \textit{dynamical quantum effects in materials}, concerned with driven and out-of-equilibrium material systems.}

\subsection{Quantum chemistry}
\label{sec:quantchem}

\begin{figure}[t]
    \centering
    \includegraphics[width=0.75\columnwidth]{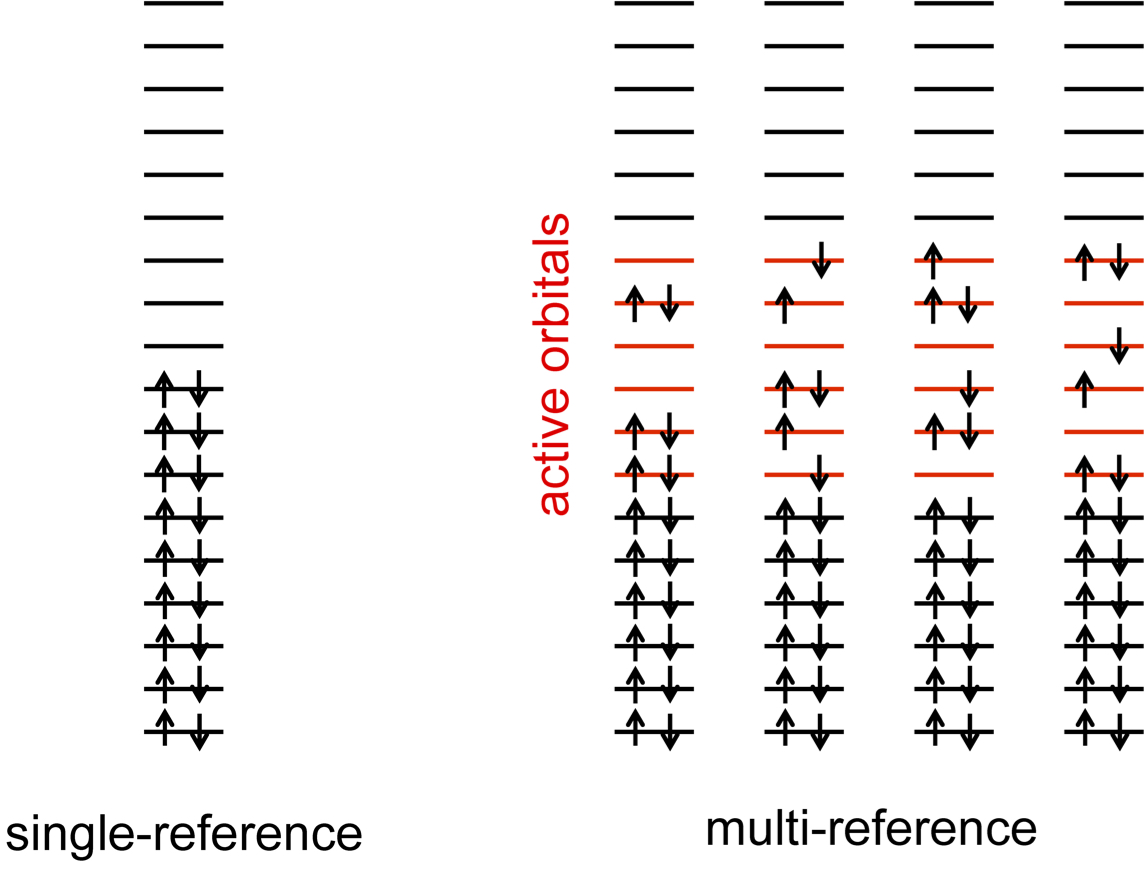}
    \caption{Single-reference (left) and multi-reference (right) wavefunctions. The former is qualitatively described by a single Slater determinant, the latter by a linear combination of a potentially large number of Slater determinants. Often
    such determinants correspond to different configurations of electrons in an ``active space'' of orbitals.}
    \label{fig:srmr}
\end{figure}

Quantum chemistry is concerned with determining the {low-lying eigenstates of the electronic Hamiltonian}
of a molecule. The eigenstates are determined for fixed sets of nuclear positions, i.e. within the Born-Oppenheimer approximation.
Determining the main features of the resulting potential energy surface, i.e. the electronic energy as a function of nuclear positions, its minima and saddle points, is key to understanding chemical reactivity, product distributions, and reaction rates.

There exists a wide range of quantum chemical methods with different accuracy and speed tradeoffs, ranging from density functional methods that
routinely treat a few \revision{thousand}  atoms or more on modern cluster resources~\cite{jones2015density}, to high-level many-electron wavefunction methods, such as coupled cluster theory,
that can attain chemical accuracy of 1 kcal/mol and better, on systems of tens of atoms~\cite{bartlett2007coupled,sparta2014chemical}. However,
most methods in quantum chemistry are most
accurate for problems where there is a dominant electronic configuration, a subset of the quantum chemistry problem
known as the \textit{single-reference problem}.
Single-reference quantum chemistry is found in the ground-states of many simple molecules (e.g. hydrocarbons), but in many molecular excited states,
at stretched bond geometries, and in transition metal chemistry,  multiple electronic configurations
can come into play, which is referred to as \textit{multi-reference quantum chemistry}. \revision{The distinction between single- and multi-reference problems in quantum chemistry is sketched in Figure~\ref{fig:srmr}.}
Despite much work (and progress) in extending quantum chemistry to multi-reference situations,
the attainable accuracy in molecules with more than a few atoms is significantly lower than
in the single-reference case. Some examples of multi-reference quantum chemical problems include:

\begin{figure*}
    \centering
    \includegraphics[width=0.75\textwidth]{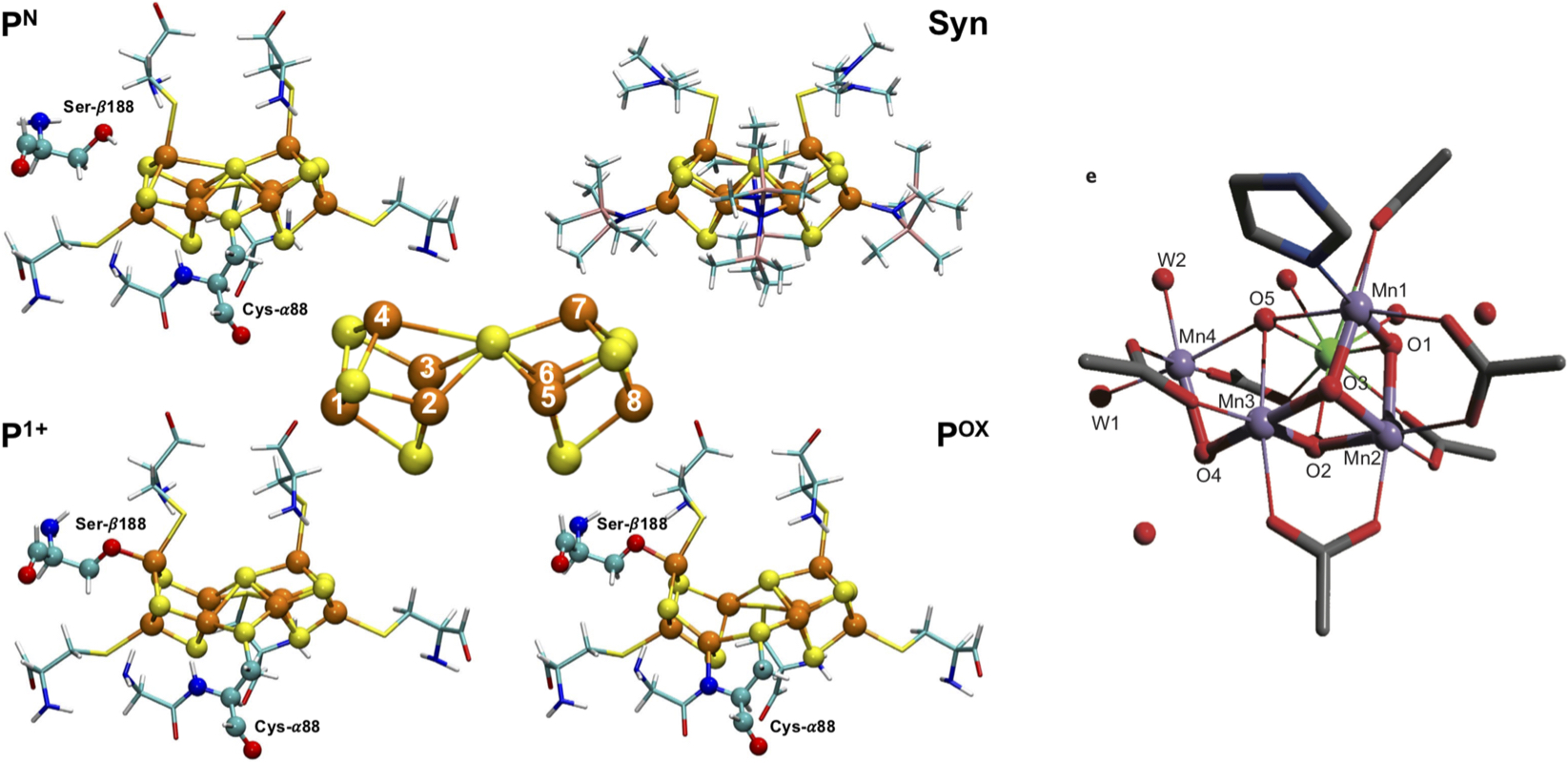}
    \caption{Left: Iron-sulfur clusters associated with different oxidation states ($\mathsf{P^{N}}$, $\mathsf{Syn}$ (a synthetic model of $\mathsf{P^N}$), $\mathsf{P^{1+}}$ and $\mathsf{P^{ox}}$)     of the $\mathsf{P}$ cluster of the nitrogenase enzyme, from \revision{Ref.~}\cite{li2019electronic}.
    Right: Mn$_4$Ca core of the oxygen evolving complex of photosystem II, from Ref.~\cite{kurashige2013entangled}.}
    \label{fig:enzymes}
\end{figure*}

  \begin{itemize}
  \item \textit{The chemistry of {enzyme active sites}}. Such active sites can involve multiple coupled transition metals,
    famous examples being the 4 manganese ions in the oxygen evolving complex~\cite{cady2008functional}, or the 8 transition metals in the iron-sulfur clusters of 
    nitrogenase~\cite{noodleman2002insights,reiher2017elucidating}, shown in \revision{ Figure~}\ref{fig:enzymes}. They present some of the most complicated multi-reference quantum chemistry problems in the
    biological world. Combined theoretical and experimental studies, primarily at the level
    of density functional theory, have proven successful in unravelling many structural and electronic features of such enzyme active sites~\cite{siegbahn2009structures,lancaster2011x,batool2019magnetic}.
    However, a detailed understanding of the interplay between spin-coupling and delocalization between metals, 
    which requires true multi-reference quantum chemistry and is needed to interpret aspects of experimental spectroscopy, 
    remains in its infancy~\cite{sharma2014low,kurashige2013entangled,li2019electronic,li2019electronic,chilkuri2019ligand,cao2018protonation}.
  \item \textit{Transition metal nanocatalysts and surface catalysts}. Similarly to enzyme active sites, simulating the mechanism of action
    of synthetic heterogeneous catalysts remains a major challenge. While density functional theory has been widely employed,
    predictions of even basic quantities such as the adsorption energy of small molecules are unreliable~\cite{schimka2010accurate,capdevila2016performance}. While not all such
    problems are expected to be multi-reference in character, even the single-reference modeling of such chemistry, at a level
    significantly beyond density functional theory, is currently challenging or impossible. In addition, multi-reference effects are
    expected to play a role in certain catalysts, such as transition metal oxides, or at intermediate geometries in reaction pathways~\cite{norskov2011density,schimka2010accurate,wodtke2016electronically,norskov2009towards}. 
  \item \textit{Light harvesting and the vision process}. The photochemistry of conjugated organic molecules is the means by which
    \revision{nature} interacts with light. Some prominent examples of such natural conjugated systems include the carotenoid and chlorophyll pigments
    in the light-harvesting complex of plants~\cite{cheng2009dynamics,polivka2004ultrafast}, as well as the rhodopsin system associated with vision
\cite{hahn2000quantum,andruniow2004structure}.
 While describing the interaction with light is
 not purely a question of electronic structure, as it involves the quantum dynamical evolution of quantum states, the quantum chemical questions
    revolve around the potential energy surfaces of the ground- and excited-states, and the influence of the environment
    on the spectrum~\cite{segatta2019quantum}. These questions are currently challenging due to the size of the systems involved as well as the varying
    degree of single- and multi-reference character in many of the conjugated excited states~\cite{tavan1987electronic,hu2015excited,parusel2000theoretical}.
  \end{itemize}
The basic quantum simulation problem is the ground-state (or low-energy eigenstate) problem for the electronic Hamiltonians, and
the basic metric is whether ground-state or low-energy eigenstate quantum algorithms yield more accurate energies (for comparable 
computational resources) than the best classical algorithms, for the problem sizes of interest. Proof-of-principle demonstrations could 
be carried out in simplified models of the above problems (e.g. in small active spaces of orbitals). However, to make real progress, 
one should also consider more quantitative models, which requires treating a large number of electrons in a large number of orbitals; 
at minimum, tens of electrons in hundreds of orbitals.
This poses new challenges for ground-state algorithms, and raises questions of how best to {represent and encode the resulting 
Hamiltonians and states}. In addition, there are many aspects of the chemical problems beyond the modeling of the electronic wavefunctions,
for example, to treat {environmental, solvent, and dynamical effects}~\cite{tomasi2005quantum}. 
\revision{It is natural to handle these by interfacing the quantum simulation with other classical simulation methods,
  for example, in molecular dynamics, where Newton's equation could be integrated using forces determined
  from the quantum simulation~\cite{carparrinello}, in QM/MM (quantum mechanics/molecular mechanics) models, where part of the environment
  would be modeled by classical charges and force-fields~\cite{qmmm}, and in simulations using implicit continuum models of solvation~\cite{cosmo}}.
Finally, although the above examples have focused on multireference and strongly correlated quantum chemistry, weakly correlated chemistry itself becomes hard to model 
classically when the number of degrees of freedom is very large~\cite{vandevondele2012linear}. These may also be interesting 
to target with quantum algorithms when sufficiently large quantum machines are available.

\subsection{Quantum molecular spectroscopy}
\label{sec:qms}

High-resolution gas-phase rovibrational spectroscopy provides an \revision{extremely precise {experimental probe of molecular structure}~\cite{mills1972molecular,watson1977vibrational,papouvsek1982molecular,wang2013anharmonic}.}
Such spectroscopy is important not only for the fundamental understanding of small molecules and the quantum control of atomic and molecular states,
but also to provide insight into the basic chemical processes and species involved, for example, in atmospheric chemistry~\cite{vaida2008spectroscopy} and in astrochemistry~\cite{heiter2015atomic}.
In larger molecules, with more than a few atoms, even the low-energy rovibrational spectrum
contains many peaks which cannot be interpreted without theoretical simulation~\cite{barone2015quantum}.
The theoretical goal is to compute the eigenstates of the \textit{nuclear Schr\"odinger equation}~\cite{bowman2008variational}.
However, unlike the electronic structure problem, there are several challenges even in setting up the best form of
the nuclear Schr\"odinger  equation to solve.
The first is that the nuclear Hamiltonian (in particular,
the nuclear-nuclear interactions) are not known
a priori because the interactions are mediated by the electrons. This nuclear potential energy term must instead be determined from
quantum chemical calculations at fixed nuclear geometries and then fitted to an appropriate functional form; this requires a large number
of high accuracy quantum chemistry calculations. The second is that nuclear ro-vibrational motion  is often far from harmonic and
not well approximated by simple mean-field theories, unlike many electronic structure problems.
Thus there is a need for a proper choice of curvilinear nuclear coordinates that decreases coupling in the nuclear potential energy (e.g. in
a harmonic system, normal modes are such a choice of coordinates) while retaining 
a simple form for the kinetic energy operator, and which also exposes the symmetry of the molecular system~\cite{sadri2012numeric,tew2003internal}. 

Once the nuclear Schr\"odinger equation has been properly formulated, one then faces the problem of representing the eigenstates.
While methods such as diffusion Monte Carlo have made progress on vibrational ground-states \cite{viel2017zeropoint}, spectroscopy involves
transitions to excited states. In this setting, tensor factorization \cite{beck2001multiconfiguration,wodraszka2015chplus} and other approaches \cite{ove2007vibrational,ove2012selected} have been explored
to approximate the rovibrational wavefunctions~\cite{wang2002new}. However, the high dimensionality and spectral congestion, requiring resolution between peaks on the scale of 1 wavenumber, proves extremely challenging~\cite{wang2008vibrational,wodraszka2015chplus}. Some famous examples include:

\begin{figure}
    \centering
    \includegraphics[width=0.75\columnwidth]{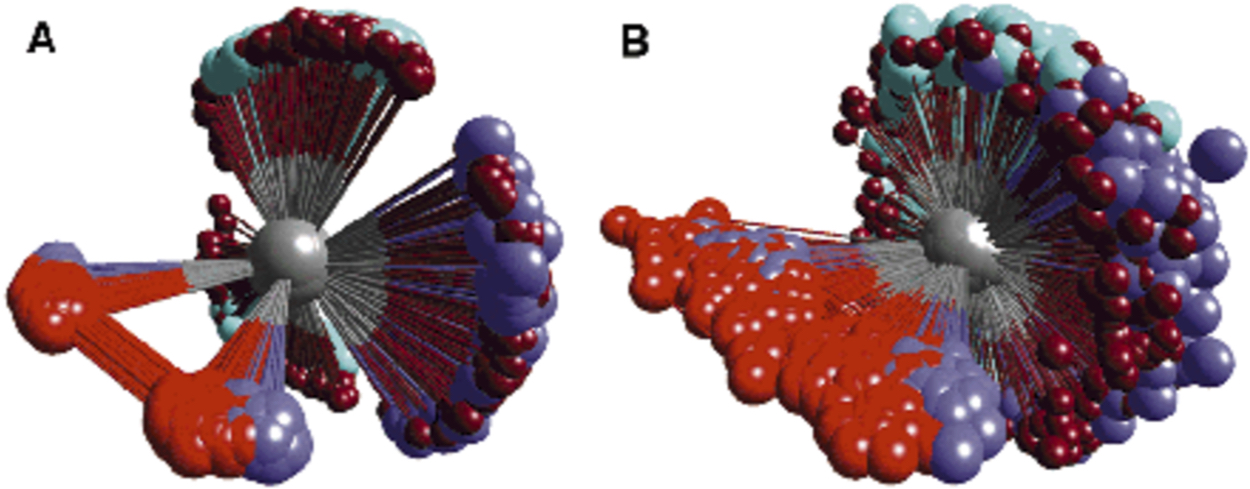}
    \caption{Dominant families of configurations of fluxional CH$_5^+$. From Ref.~\cite{marx1999ch5+}.}
    \label{fig:ch5plus}
\end{figure}

\begin{itemize}
\item \textit{Spectra of floppy molecules}. Floppy molecules are by their nature very anharmonic
  and thus far from a simple vibrational description. CH$_5^+$ is a prototypical floppy molecule (see \revision{Figure~}\ref{fig:ch5plus}):
  the five hydrogen atoms move around the central carbon and the molecule
  has almost no structure in the traditional sense~\cite{oka2015taming,white1999ch5+,huang2006quantum,wang2016calculated,wodraszka2015chplus}.
\item \textit{Hydrogen bonded clusters}.   Another vibrational problem with large anharmonicity is found
  in hydrogen bonded clusters, such as in the spectroscopy of water molecules and protonated water clusters~\cite{vendrell2009full,yu2019classical}. The hydrogen bond network is fluid
  and even small clusters can transition between many different
  minima on the potential energy surface~\cite{liu1996characterization}. Resolving the peaks and tunnelling splittings is important for interpreting
  water spectra in the atmosphere, as well as in understanding reaction mechanisms in water.
 Further, the spectroscopy of molecules with intermolecular hydrogen bonds, such as the
  malonaldehyde molecule, has also posed long-standing challenges for the field~\cite{wang2008full,coutinho2004ground,thorsten2011intramolecular,schroder2011theoretical,arias1997proton}.
\end{itemize}
From a quantum algorithms perspective, although there are similarities with the quantum chemistry problem (in particular one is 
interested in low-energy eigenstates), there are  significant differences. One important difference is that the Hamiltonian
is no longer of simple two-particle form due to the effective nuclear-nuclear interaction
and typically includes important three- and four-mode terms.
Also, one is often interested in an order of magnitude more states (e.g. hundreds of excited states) than in the electronic
structure problem. All these features are sufficiently distinct from the usual quantum chemical scenarios that quantum
algorithms are likely to require additional innovation to be useful in the nuclear problem. Some
steps in this direction have recently appeared~\cite{sawaya2019vibrations,mcardle2019vibrations,wang2019bosonic}.
One simplification is that many nuclei are distinguishable avoiding the need to consider indistinguishable particles. 
The lack of a good mean-field starting point together with the various technical complications means that one can find
relatively small systems (in terms of the Hilbert space size) where classical methods already have trouble; for example, a
minimal quantum model of the CH$_5^+$ molecule can be formulated as a 12 dimensional problem
with 10 basis functions per mode~\cite{wang2016calculated}.

\subsection{Chemical quantum dynamics}
\label{sec:cqd}

Chemical quantum dynamics is another important target for quantum simulation~\cite{child1996molecular,nitzan2006chemical,goldfield2007quantum,clary2008quantum,greene2017ttsoft}. This field is concerned with
modeling time-dependent electronic and nuclear quantum effects in molecules (as distinct from
computing the time-independent electronic or nuclear eigenstates in quantum chemistry and quantum molecular spectroscopy).
\textit{Quantum molecular dynamics} is primarily concerned with the nuclear motion and describes the rates of chemical processes as well
as the dynamical interaction of molecules with light, as involved in spectroscopy and quantum control.
However, with the development of short X-ray pulses direct experimental access to electron dynamics in molecules is now also available~\cite{nisoli2017attosecond}.

Currently these dynamical simulations are challenging. For example, nuclear motion is poorly described by mean-field theory and 
the classical limit is  often a better starting point, but offers no zeroth order description of quantum effects.
Thus classical simulations of quantum dynamics either invoke methods
based on the classical limit that scale to large systems but which are difficult to systematically improve (such as approximate path integral methods~\cite{craig2004quantum}, \cite{cao1994formulation}),
or methods which model the wave-function dynamics or the path integral accurately for 
a small number of degrees of freedom~\cite{greene2017ttsoft,wang2003multilayer}, but which are not scalable due to dimensionality or the dynamical sign problem~\cite{makarov1994path}.

A subfield of quantum molecular dynamics, but one of important chemical interest, is the description of \textit{non-adiabatic quantum effects} \cite{nitzan2006chemical}.
At nuclear configurations where different electronic surfaces approach each other, the Born-Oppenheimer approximation
can break down and the quantum behavior of the nuclei, coupled indirectly via the electrons, is enhanced.
The associated quantum non-adiabatic effects
govern non-radiative energy relaxation via the crossing of electronic surfaces (so-called {conical intersections}) and are thus central 
in describing energy transfer. The faithful description of non-adiabatic quantum effects requires the simultaneous treatment of quantum 
electrons and quantum nuclei. The complexity of this problem together with the often large system sizes where non-adiabatic effects are of 
interest means that current classical methods rely on simple heuristic approximations, such as surface hopping~\cite{barbatti2011nonadiabatic}, for which a rigorous 
quantum formulation is lacking. Examples of relevant chemical problems in the area of chemical quantum dynamics include:

\begin{itemize}

\item \textit{Proton coupled electron transfer} (PCET) \cite{huynh2007proton}. 
PCET is known to be an important mechanism in catalysis and energy storage: electrons are transferred at lower overpotentials when thermodynamically coupled to proton transfer. Examples range from homogeneous catalysts \cite{roc2015intramolecular,ge2017interfacial,wang2010study,amin2013electrostatic}
and heterogeneous electrocatalysts
\cite{chen2016ultrafast,liao2014electrochemical}
to enzymes that perform PCET, such as soybean lipoxygenase (that catalyzes the oxidation of unsaturated fatty acids) \cite{hatcher2004proton}, photosystem II (in the tyrosine oxidation step) \cite{sjodin2000proton,sproviero2008model}
and the redox-leveling mechanism of catalytic water oxidation
\cite{liao2014electrochemical}.
While semiclassical predictions are often good enough for describing electron transfer, the quantum nature of molecular dynamics is paramount with PCET because of the quantum nature of the proton\cite{hatcher2004proton,carra2003proton} as evidenced in kinetic isotope effects (referring
to the ratio of the proton to deuteron reaction rates) which can be very large \cite{hatcher2004proton}. Thus, classical mechanics is not sufficient, and quantum simulations of PCET would be extremely helpful for making predictions in large (especially biological) systems.

\item  \textit{Vibrational dynamics in complex environments}.  For many systems of interest, vibrational spectroscopy is the key tool available for characterization.   There is overlap with the problems in \revision{Subsection}~\ref{sec:qms} but the questions here focus  on larger scale systems and condensed phase problems,
  where the line-shapes as well as frequencies are important, and the system size  limits the use of fully quantum formalisms.
Indeed, from the librations of water to the high frequency motion of C-H bonds, much of physical chemistry uses nuclear vibrational frequencies to characterize complex systems.

Nevertheless, because of computational limitations, the standard approach today for modeling vibrational dynamics in large complex environments
is to invoke a very old flavor of theory: Kubo theory \cite{nitzan2006chemical}.  In this case, one focuses on energy gaps and uses a semiclassical expansion of the line-shape. However, even for the energy gaps one relies on diagonalization
of the quantum subsystem and this 
limits the problems that are accessible; systems with
many interacting quantum states -- for instance, very large H/J molecular aggregates \cite{riehl1986circularly} -- 
are difficult or out of reach.

\item  \textit{Plasmonic chemistry}. A recent development is the  possibility of using metal particles (with large cross sections) as a tool to  absorb light and, with the resulting plasmonic excitations, initiate ``plasmonic chemistry''~\cite{cortes2018activating}.  Already, there are a few well-studied examples in the literature of hot plasmonic chemistry, such as the plasmon induced dissociation of hydrogen molecules on nanoparticles~\cite{mukherjee2012hot}. To better understand this arena,
however, one of the key questions is: how do we characterize plasmonic excitations?  While classical descriptions of plasmons are easy to obtain,
quantum descriptions are necessary  to model quantum processes, e.g. electron transfer. Covering
the disparate length scales of plasmonic excitations and the chemical process remains  extremely difficult. 
\end{itemize}

While quantum dynamics is in principle an ideal simulation problem for a quantum computer,
the quantum simulation of quantum molecular dynamics entails several practical challenges. Some of this
may be viewed as an issue of representation. As in
the problem of quantum molecular spectroscopy, the nuclear quantum Hamiltonian
contains complicated interactions which must be tabulated or calculated on the fly~\cite{kassal2008polynomial}.
In addition to this, the dynamical quantum state involves near
{continuum degrees of freedom}, posing a challenge for the standard discretizations of Hilbert space considered
in quantum algorithms. Finally, typical spectroscopic observables may be accessible to relatively simple
treatments, not requiring the full fidelity of the quantum wave-function evolution.
These technical issues mean that implementing quantum molecular dynamics with a quantum advantage is likely to remain
challenging in practice, despite the favorable theoretical complexity on a quantum device.

\subsection{Correlated electronic structure in materials}
\label{sec:ces}

\begin{figure}
    \centering
    \includegraphics[width=0.75\columnwidth]{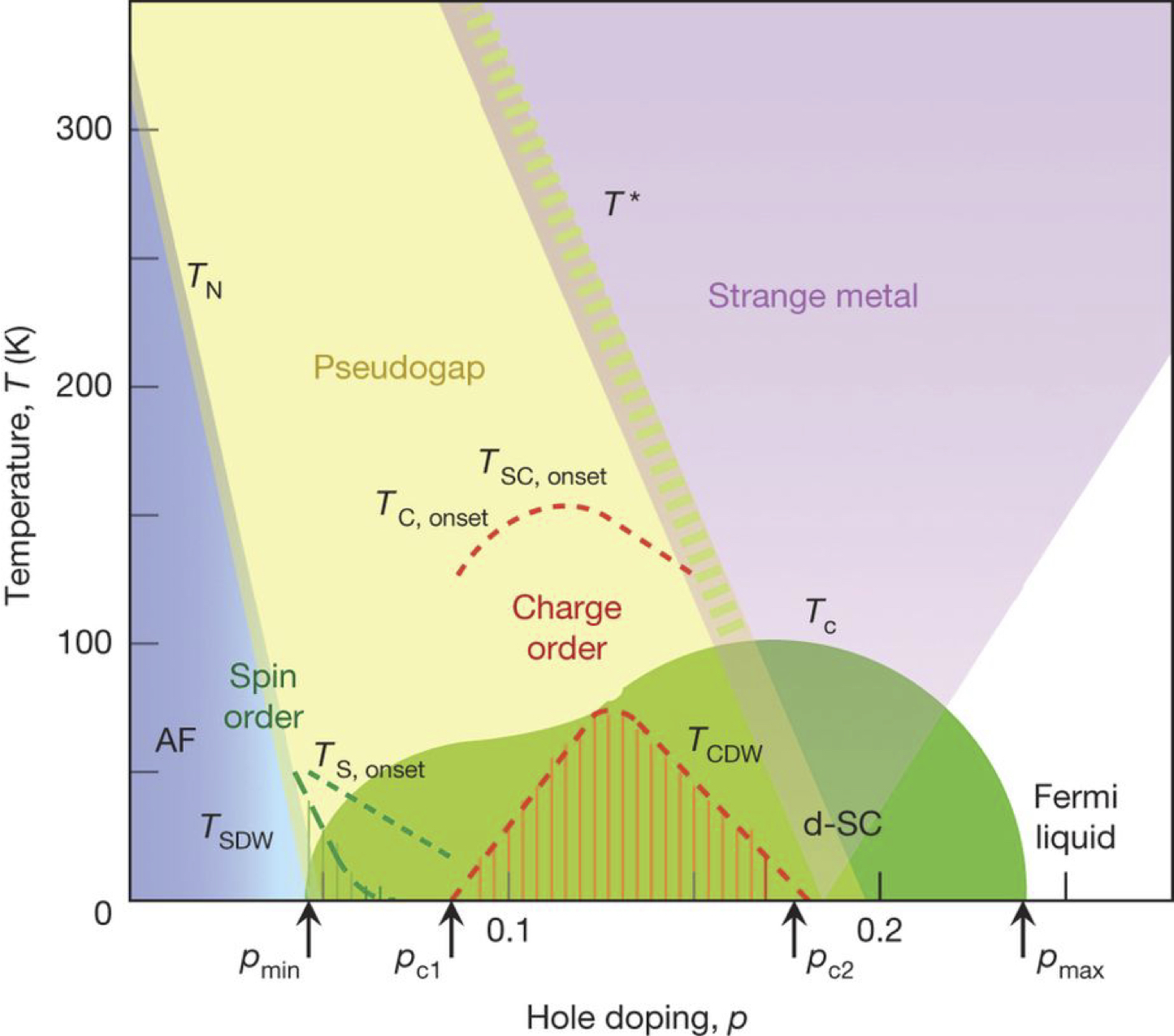}
    \caption{Qualitative phase diagram of cuprate high-temperature superconductors, which has challenged theory and simulation in condensed matter for decades. Figure adapted from Ref.~\cite{keimer2015quantum}.}
    \label{fig:high-tc-phase-diagram}
\end{figure}

The goal of electronic structure calculations in materials is to {determine their low-energy properties}. There is some overlap
in methods and ideas between the materials electronic structure problem and the problem of quantum chemistry.
When electron-electron interactions are weak, the low energy material properties can normally be described by computing the band structure using \textit{density-functional theory} and one of the many popular density functionals, such as the local density approximation (LDA) or generalized gradient approximations (GGA) \cite{GGA1,GGA2,GGA3}. However, in some materials, commonly referred to as \textit{strongly correlated}, the electron-electron interactions fundamentally alter the behavior and such an effective non-interacting description is no longer appropriate. A paradigmatic example are Mott insulators, which appear as conductors in band structure theory but in fact become insulating due to electron-electron interactions. While the mechanism behind the insulating behavior of Mott insulators is well-understood, for many other phenomena in strongly correlated systems {the underlying microscopic mechanism is not fully known}, let alone a quantitative and predictive theory of the associated physics with material specificity. Some famous examples of such problems include:
\begin{itemize}

\item Originally discovered in 1986, \textit{high-temperature superconductivity} \revision{has eluded a theoretical explanation to date \cite{leggettNP}}. The experimental phase diagram, which is sketched in Figure~\ref{fig:high-tc-phase-diagram}, is accurately characterized experimentally across several materials. While the general properties of the superconducting phase itself are relatively well characterized, the mechanism driving superconductivity
is not yet fully elucidated. Also, two nearby regimes, the \textit{pseudogap} and \textit{strange metal} phase, continue to puzzle theorists~\cite{keimer2015quantum}. In both cases, their nature as well as their precise relation to the superconducting phase are not understood. The strange metal phase (also known as non-Fermi liquid) exhibits behavior inconsistent with a simple weakly interacting metal even at high energies, and has motivated a whole area of research on exotic metallic systems~\cite{lee2018recent}. Meanwhile, the pseudogap phase exhibits several competing ordering tendencies~\cite{fradkin2015intertwined}, which are extremely challenging to resolve in numerical methods because most methods naturally favor a particular ordering pattern, thus making it challenging to disentangle physical effects from method biases.

\item The non-Fermi liquid behavior exemplified in the strange-metal phase of cuprates appears also in other classes of materials, such as \textit{heavy fermion compounds} and \textit{fermionic systems near criticality}~\cite{coleman2007heavy,si2010heavy}. Often, these systems are amenable to classical simulation only at special points where quantum Monte Carlo methods do not suffer from the infamous sign problem~\cite{li2019sign}.

\item Two-dimensional systems have long been of central interest in strongly correlated physics. Many material systems realize effectively two-dimensional physics, including two-dimensional electron gases in semiconductor heterostructures, where the \textit{integer and fractional quantum Hall effect} were first discovered~\cite{tong2016lectures}, layered materials (including cuprate high temperature superconductors), graphene~\cite{castroneto2009} and transition-metal dichalcogenides (TMDs). More recently, it has been found that so-called \textit{Moir\'{e} materials} exhibit rich phase diagrams due to strong interactions, including exotic superconductivity~\cite{cao2018unconventional} and exotic topologically non-trivial phases. A paradigmatic example is twisted bilayer graphene~\cite{macdonald2019trend}, which consists of two graphene layers that are slightly twisted with respect to each other. This leads to a Moir\'{e} pattern with a very large unit cell, which effectively quenches the kinetic energy (i.e., leads to almost flat bands) and drastically enhances the effect of Coulomb interaction.

\item \textit{Frustrated spin systems} have long been an important topic especially for numerical simulations in condensed-matter physics. These systems potentially realize a host of high non-trivial phases, in particular topological and gapless spin liquids~\cite{balents2010spin}. They have historically been the testbed for computational methods such as tensor networks and variational methods. As such, they appear as good test cases also for quantum simulations. Furthermore, recent developments in particular in materials with strong spin-orbit coupling have opened the door on a variety of new materials that may exhibit exotic topological phases, and in particular realize a non-Abelian spin liquid~\cite{kitaev2006anyons,kasahara2018unusual}.

\end{itemize}

While many methods have been developed to accurately include electron-electron interactions, their scope generally remains limited. For example, {tensor network methods} have revolutionized the study of one- and to a limited extent two-dimensional effective models for magnets and itinerant fermions. However, these methods so far have not been successfully applied to more realistic models, and in particular in three dimensions. On the other hand, quantum embedding methods such as the {dynamical mean-field theory} (DMFT) and its many cousins can capture interaction effects in three-dimensional systems, including multi-band systems. However, they require some approximations to the correlations of the state; for example, in its simplest form, DMFT disregards momentum dependence of the electron self-energy. While many improved variants of these embedding methods exist, their accuracy is often difficult to control, and so far
they have not been applied to realistic models without further approximations. Finally, {quantum Monte Carlo} methods have been extremely successful for bosonic systems and unfrustrated spin systems, but the sign problem hinders their application to frustrated or fermionic systems (away from special points) without other uncontrolled approximations.

From the perspective of quantum algorithms, the materials electronic structure problem is both simpler
and more difficult than the quantum chemistry problem. Some ways in which it is simpler include the fact that often very simple Hamiltonians
describe the main physics, as well as the potential presence of translational invariance. A major way in which it is more complicated is
the fact that one needs to treat systems approaching the thermodynamic limit, which involves a very large number of degrees of freedom.
This not only increases the number of qubits required but also heavily impacts the circuit depth of algorithms, such as state preparation.
The thermodynamic limit can also lead to small energy scales for excitations and energy differences between competing phases. For these reasons,
it remains to be understood whether the quantum algorithms of relevance to quantum chemistry are those of relevance to materials electronic structure.

\subsection{Dynamical quantum effects in materials}
\label{sec:dqem}

\begin{figure}[t]
    \centering
    \includegraphics[width=0.75\columnwidth]{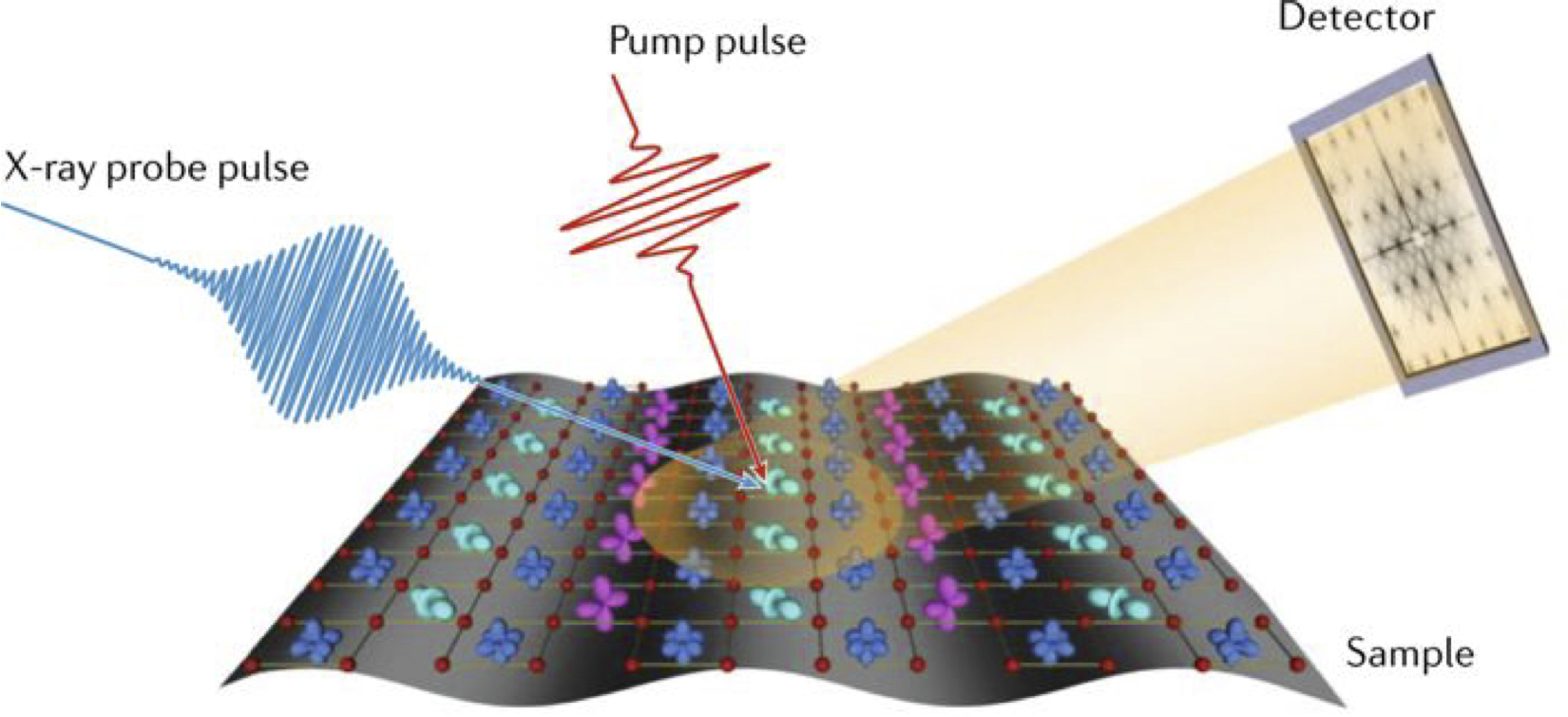}
    \caption{Illustration of pump-probe spectroscopy using X-ray probes. Figure adapted from Ref.~\cite{buzzi2018probing}.}
    \label{fig:pump-probe}
\end{figure}

Many experiments on condensed-matter systems do not probe the equilibrium properties of the system, but rather dynamical properties. For example, the main workhorse of mesoscopic quantum physics is \textit{electron transport}, i.e. the response of the system when it is coupled to electron reservoirs and a voltage is applied~\cite{nazarov2009quantum}. Likewise, material properties are often probed through scattering experiments, such as neutron scattering or angle-resolved photoemission spectroscopy (ARPES)~\cite{Damascelli_2004}, which probe dynamical properties such as the {structure factor or spectral function}. Going beyond spectral properties, the \textit{non-equilibrium real-time dynamics} of quantum systems has increasingly come into focus, both because of experiments that can probe quantum dynamics at atomic scales, \revision{and because of fundamental interest in studying the equilibration of quantum systems,
which serves as a bridge between the theories of quantum mechanics and statistical mechanics, as discussed below}. Experimental
setups that can probe ultra-fast dynamics in materials include, for example, free-electron lasers~\cite{patterson2010coherent,weathersby2015mega} as well as other pulsed laser systems. These allow the application of experimental techniques, such as pump-probe spectroscopy~\cite{fischer2016invited,buzzi2018probing}, \revision{illustrated in Figure~\ref{fig:pump-probe},} to provide novel insights into the behavior of correlated quantum systems.

\begin{figure}
    \centering
    \includegraphics[width=0.75\columnwidth]{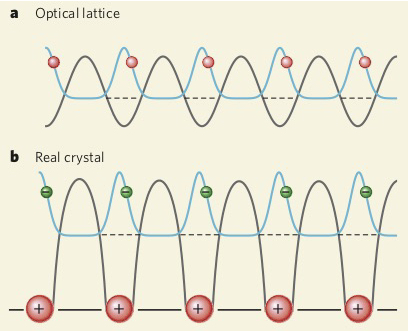}
    \caption{Cold atomic gases are controllable quantum systems that can be arranged in a lattice (a). They can be tuned to provide analog simulators of real crystal potentials (b). From Ref.~\cite{greiner2008condensed}.}
    \label{fig:optical_lattice}
\end{figure}

On the other hand, \textit{cold atomic gases}~\cite{bloch2008many} provide a highly controllable environment that allows systematic exploration of quantum dynamics even in the strongly interacting regime~\cite{langen2015ultracold}, see \revision{Figure~\ref{fig:optical_lattice}}. A key advantage is that one can engineer the evolution of the system to closely follow a target model; this approach is also referred to as analog quantum simulation (see also \revision{Subsection}~\ref{sec:quantum-architectures}). However, classical simulation still plays a crucial role in establishing the accuracy of cold atom setups.

From a conceptual point of view, a central question has become \textit{the connection between statistical mechanics and the dynamics of closed quantum systems}. The general goal is to put quantum statistical mechanics on a solid conceptual foundation as well as understanding the cases where it does not apply, such as many-body localized systems~\cite{nandkishore2015many,dalessio2016quantum,abanin2019colloquium}.

Numerically simulating all these systems has been a severe challenge. While many approaches to quantum dynamics exist, none are universally applicable. At short times, tensor-network methods can accurately describe the dynamics. 
However, in general the computational cost grows exponentially with the desired simulation time, thus severely limiting the 
timescales that can be resolved~\cite{paeckel2019time}. Conversely, at long times, the system is often effectively described by classical dynamics controlled by 
conservation laws. However, the interesting strongly correlated behavior is generally exhibited at intermediate times, 
inaccessible to the established classical methods. For example, quantum Monte Carlo methods can scale exponentially in time even for unfrustrated 
systems. Non-equilibrium dynamical mean-field theory~\cite{aoki2014} has emerged as a powerful method especially for systems in high dimensions, but requires uncontrolled approximations (both in the setup of the method and its numerical solution). Finally, in the regime of weak interactions, time-dependent density functional theory can be used, but likewise implies uncontrolled approximations~\cite{marques2006time}.

\section{Challenges for quantum algorithms in quantum simulation}
\label{sec:3}

\subsection{Overview of algorithms}

In Section~\ref{sec:intro} we described three quantum problems that lie at the heart of chemistry
and materials physics: the problem of \textit{quantum dynamics}, representative of computational tasks that can be efficiently
tackled by a quantum computer, aside from the initial state preparation; \textit{quantum ground- and low-energy state determination}, central
to quantum chemistry, condensed phase electronic structure,
and quantum molecular spectroscopy; and {thermal averages} or \textit{quantum statistical mechanics}, to describe
finite-temperature chemistry and physics. In next sections, we survey the current status and
theoretical and practical challenges to implement quantum algorithms for these problems, and some future directions. 

Regardless of the problem we are studying, the first step in a quantum simulation
is to choose a \textit{representation} for the Hamiltonian and the states. Thus, in \revision{Subsection~}\ref{sec:map2qubits} we first
examine the possibilities for different qubit representations, and open questions in that area.

Quantum ground-state algorithms fall into several classes. \textit{Quantum phase estimation} is a
direct route to (near exact) eigenstate determination, but has been challenging to implement so far~\cite{pea1,pea2,o2016scalable}.
A complementary technique is to prepare the exact ground-state via a prescribed ``exact'' evolution path,
either in {real time} (\textit{adiabatic state preparation}), or in {imaginary time} (\textit{quantum imaginary time evolution}) \cite{Messiah1962,veis2014adiabatic,Motta2019}. 
On the other hand, \textit{variational quantum algorithms} provide the possibility to introduce approximations with adjustable circuit depth,
which are determined via a non-linear optimization, usually implemented in a hybrid-quantum classical loop \cite{peruzzo2014variational,mcclean2016theory,McArdle_2019}.
The variety of algorithms currently explored for quantum ground- (and low-lying excited state) determination are discussed in 
\revision{Subsection~}\ref{sec:quantum_ground_state}.

While polynomial time algorithms for \textit{quantum time evolution} have been known for some time \cite{lloyd1996universal},
algorithms with optimal asymptotic complexity (linear in time) as well as favorable scaling with error and low prefactors, remain an area of active research \cite{low2019hamiltonian-2,childs2018toward}. Also, much work remains to be done to optimize Hamiltonian simulation algorithms for the Hamiltonians of interest in chemistry and materials
science. Quantum {time evolution} is a fundamental building block in many quantum algorithms, such as phase estimation \cite{aspuru2005simulated}.
The current status of quantum time evolution algorithms is summarized in \revision{Subsection~}\ref{sec:quantum_time_evolution}.

How best to simulate {thermal states} in chemical and materials science applications remains an open question. A wide variety of techniques
have been discussed, ranging from eigenstate thermalization, to state preparation methods, to hybrid quantum-classical algorithms, though
few have been implemented. The current status of thermal state methods, the prospects for implementing them, and other open questions are
discussed in \revision{Subsection~}\ref{sec:quantum_thermal_states}.

Many quantum algorithms involve interfacing with classical data and classical algorithms. This can be \revision{leveraged} to \revision{incorporate classical optimization strategies in the structure of the method}, as in variational quantum algorithms \cite{mcclean2015theory}. Another reason is to enable a multi-level/multi-scale
representation of the problem. Quantum embedding provides a framework for such multi-scale quantum/classical hybrids, with the
quantum representation of a subsystem coupled either to a classical environment, or another quantum representation via the
exchange of classical data. We discuss the current status of \textit{hybrid quantum-classical algorithms and quantum embedding} in particular
in \revision{Subsection~}\ref{sec:quantum_classical}.

An important consideration when developing improved quantum algorithms for real chemical and materials science problems
is to establish \textit{benchmark systems and results}, from the best available classical simulation data. The possibilities and prospects for
such benchmarks are discussed in \revision{Section~}\ref{sec:benchmarks}.

\revision{Finally, we note that the organization we have adopted has been  driven by theoretical and technical distinctions and thus
  we make no attempt to describe the historical development and organization of the field. Readers interested in the earliest papers in quantum simulation,
  which often adopt a different language from more modern discussions, may wish to refer to references~\cite{feynman1982simulating,meyer1996quantum,
boghosian1998quantum, wiesner1996simulations,
Lloyd96, abrams1997simulation, Terhal_PRA_2000,
kitaev2002classical,
somma2003quantum,
aspuru2005simulated}.}

\subsection{Qubit representation of many-body systems}
\label{sec:map2qubits}

Many-body systems in chemistry and materials physics are
composed of interacting electrons and  atomic nuclei. An exact quantum mechanical treatment
involves continuous variables such as the particles' positions and momenta.
To simulate such systems on a digital computer (either quantum or classical),
the infinite-dimensional Hilbert space of a many-body system has
to be truncated.  

The most direct route is to define a finite set of \textit{basis functions} and
then to project the exact many-body Hamiltonian onto the chosen basis.
The resulting discretized system is then expressed in terms of qubits. Depending
on the problem, the Hamiltonian of interest may be different, e.g. in electronic structure it is the electronic Hamiltonian,
while in molecular vibrational problems, it is the nuclear Hamiltonian. Alternatively, one can write down a simple
form of the Hamiltonian a priori that contains the main interactions (a model Hamiltonian) with adjustable parameters. This latter
approach is particularly popular in condensed matter applications.
Finally, depending on the particles involved it may also be necessary to account for their fermionic or bosonic nature, in which case a suitable
encoding of the statistics is required.

A choice of a good representation is important as it may affect the simulation cost dramatically.
In this section we briefly summarize known methods for
the qubit representation of many-body systems, discuss their relative merits, and outline important directions for future research. 

\subsubsection{Ab initio electronic structure qubit representations}

The main objective of electronic structure in chemistry and physics to understand the low-energy
properties of the electronic structure Hamiltonian
that describes a system of interacting electrons moving in the 
potential created by atomic nuclei~\cite{helgaker2014molecular,martin2016interacting}, 
\begin{equation}
\begin{split}
\label{qchem}
\hat{H} &=
\hat{H}_1+\hat{H}_2 
\;,\; \\
\hat{H}_1&=\sum_{i=1}^K -\frac12 \Delta_i^2 + V(r_i)
\;,\;
\hat{H}_2 = \sum_{1\le i<j\le K} \; \frac1{|r_i - r_j|}
\;.
\end{split}
\end{equation}
Here $K$ is the number of electrons, 
 $r_i$ is the position operator of the $i$-th electron, 
 $\Delta_i$ is the corresponding Laplacian,
 and $V(r)$ is the electric potential created
by atomic nuclei at a point $r$. The term $\hat{H}_1$ includes the kinetic and the potential energy of
non-interacting electrons while $\hat{H}_2$ represents the Coulomb repulsion. 
Here we  ignore relativistic effects
and  employ the standard Born-Oppenheimer approximation to solve the electronic Hamiltonian
for fixed nuclei positions. 

Each electron is described quantum mechanically by its position $r_i\in \RR^3$ 
and spin $\omega_i\in \{\uparrow,\downarrow\}$. Accordingly, 
a quantum state of  $K$ electrons can be specified by a wave function 
$\Psi(x_1,\ldots,x_K)$, where $x_i=(r_i,\omega_i)$.
The wave function must obey Fermi statistics, that is,
$\Psi(x_1,\ldots,x_K)$ must be anti-symmetric under exchanging
any pair of coordinates $x_i$ and $x_j$.

The first step of any quantum electronic structure simulation algorithm is to approximate
the electronic Hamiltonian $\hat{H}$ with a simpler simulator Hamiltonian
that describes a system of interacting qubits.
This is usually achieved 
by truncating the Hilbert space of a single electron to a
finite set of basis functions $\psi_1,\ldots,\psi_N$ known as (spin) orbitals. 
For example, each  orbital  could be a linear combination of atom-centered Gaussian functions
with a fixed spin orientation. 

Electronic structure simulation algorithms based on the \textit{first quantization} method~\cite{abrams1997simulation,babbush2017exponentially,kivlichan2017bounding}
describe a system of $K$ electrons
using the  Configuration Interaction (CI) space (in classical simulations, this would be called the Full Configuration Interaction space).
It is convenient to choose Slater determinants as a basis for this space. 
\revision{Slater determinants are many-body wavefunctions  defined as
  \begin{align}
    \Psi(x_1, \ldots, x_K ) \propto \det
  \begin{bmatrix}
    \psi_1(x_1) & \psi_1(x_2) & \ldots &\psi_1(x_K) \\
    \psi_2(x_1) & \psi_2(x_2) & \ldots &\psi_2(x_K) \\
    \vdots & & &\vdots \\
    \psi_K(x_1) & \psi_K(x_2) & \ldots &\psi_K(x_K)
  \end{bmatrix}
  \end{align}
  where $\{ \psi_i(x_i) \}_{i=1}^N$ are a set of orthonormal spin-orbitals.
  The set of Slater determinants spanning the CI space \revision{$\mathcal{H}_{K,N}$}
  is formed by distributing $K$ electrons over the $N$ one-electron spin-orbitals in all possible ways,
  thus it has dimension ${N \choose K}$ and can be identified with the anti-symmetric subspace of $(\CC^{N})^{\otimes K}$.}

The projection of the full electronic Hamiltonian $\hat{H}$ onto the CI space of antisymmetric functions has the form
\begin{equation}
\begin{split}
\hat{H} &= 
\sum_{i=1}^K \sum_{p,q=1}^{N} t_{pq} |p\rangle\langle q|_i \\
&+ 
\sum_{1\le i\ne j\le K}\;  \sum_{p,q,r,s=1}^{N} \; 
u_{pqrs} |p\ra\la r|_i \otimes |q\ra\la s|_j.
\end{split}
\end{equation}
Here, \revision{as above, $|p\ra \equiv |\psi_p\ra$ are orthonormal one-electron spin-orbitals}. The coefficients
$t_{p,q}$ and $u_{pqrs}$  
are known as one- and two-electron integrals. For example, 
\begin{equation}
t_{pq} = \la \psi_p| \left( -\frac{\Delta}{2} + V \right)|\psi_q\ra.
\end{equation}
Likewise, $u_{pqrs}$  is the matrix element of the Coulomb
interaction operator $1/|r_1-r_2|$ between anti-symmetrized versions
of the states $\psi_p\otimes \psi_q$ and $\psi_r\otimes \psi_s$.
 Each copy of the single-electron Hilbert space $\CC^{N}$ is then encoded by a register
of $\log_2{N}$ qubits.  This requires $n=K\log_2{N}$ qubits in total.
The CI Hamiltonian $H$  includes multi-qubit interactions among
subsets of $2\log_2{N}$ qubits.
The full Hilbert space of $n$ qubits contains many unphysical states that
do not originate from the CI space. Such states have to be removed from simulation
by enforcing the anti-symmetry condition. This can be achieved by adding
suitable energy penalty terms to the CI Hamiltonian~\cite{bravyi2017tapering}. 

An important parameter that affects the runtime of quantum simulation algorithms 
is the sparsity of the simulator Hamiltonian. A Hamiltonian $H$ is said to be 
$d$-sparse if the matrix of $H$ in the standard $n$-qubit basis
has at most $d$ non-zero elements in each row \revision{(or equivalently in each column)}.
For example, the runtime of simulation algorithms based on quantum signal
processing~\cite{low2017optimal} scales linearly with the sparsity $d$.
The CI Hamiltonian $H$ has sparsity $d\sim (KN)^2$.
Thus the first-quantization method is well-suited for high-precision
simulation of small molecules when the number of electrons $K=O(1)$ is fixed and
the number of orbitals $N$ is a large parameter. As one approaches the
continuum limit $N\to \infty$, the 
number of qubits grows only logarithmically with $N$ while the
sparsity of $H$  scales as $d\sim N^2$.

The \textit{second quantization} approach often results in a simpler simulator Hamiltonian and requires
fewer qubits, especially in the case when the filling fraction $K/N$ is not small. 
This method is particularly well suited for quantum simulation algorithms~\cite{aspuru2005simulated}
and has been experimentally demonstrated for small molecules~\cite{kandala2017hardware}.
Given a set of $N$ orbitals $\psi_1,\ldots,\psi_N$, the second-quantized simulator
Hamiltonian is 
\begin{equation}
\label{Q2}
H =\sum_{p, q=1}^{N} t_{p q} \, \hat{c}^\dag_{p} \hat{c}_{q} + 
\frac{1}{2} \sum_{p, q, r, s = 1}^{N} u_{pqrs}\, \hat{c}^\dag_{p} \hat{c}^\dag_{q} \hat{c}_{r} \hat{c}_{s},
\end{equation}
where $\hat{c}_p^\dag$ and $\hat{c}_p$ are the creation and annihilation operators for the orbital $\psi_p$.
The Hamiltonian $H$ acts on the Fock space
\revision{$\mathcal{F} = \oplus_{K=0}^{N} \mathcal{H}_{K,N}$ of an arbitrary number of fermions in $N$ spin-orbitals. The Fock space is}
spanned by $2^N$ basis vectors 
$|n_1,n_2,\ldots,n_N\ra$, where $n_p \in \{0,1\}$  is the occupation number of the orbital $\psi_p$.
The advantage of the second quantization method is 
that the Fermi statistics is automatically enforced at the operator level.
However, the number of electrons can now take arbitrary values between $0$ and $N$.
The simulation has to be restricted to the subspace with exactly $K$ occupied orbitals.
The second-quantized Hamiltonian $H$ can be written in terms of qubits
using one of the fermion-to-qubit mappings discussed in \revision{Subsection}~\ref{sec:f2q}.

Within the above outline, there are several active areas of research. For example, one may wonder if
the redundancy of the space in the first quantized representation can be reduced or completely avoided.
Other questions include the choice of basis functions and fermion-to-qubit mappings. These are discussed in the next sections.

\subsubsection{Electronic basis functions}

The discretization of the electronic Hilbert space for a quantum simulation requires balancing two concerns.
We need to represent the state with as few qubits, but also, retain maximal Hamiltonian sparsity.
These requirements only partially align with those of classical many-particle quantum simulations.
 In the classical setting, \revision{representing a quantum state at polynomial cost} is crucial due to the exponential Hilbert space, while
Hamiltonian sparsity is less so; the choice of basis functions
has historically been made so that matrix elements of the Hamiltonian (the one- and two-electron ``integrals'')
can be analytically evaluated~\cite{szabo2012modern,helgaker2014molecular}.

There are two families of basis functions in wide use in quantum chemistry and quantum materials science:
atomic orbital \textit{Gaussian bases} and \textit{plane waves}. Gaussian bases are most commonly employed
in molecular simulations due to their compactness, while plane waves are most often used in crystalline materials simulation, due
to their intrinsic periodicity and ease of regularizing the long-range contributions of the Coulomb operator (which are
conditionally convergent in an infinite system). 

\begin{figure*}
    \centering
    \includegraphics[width=0.75\textwidth]{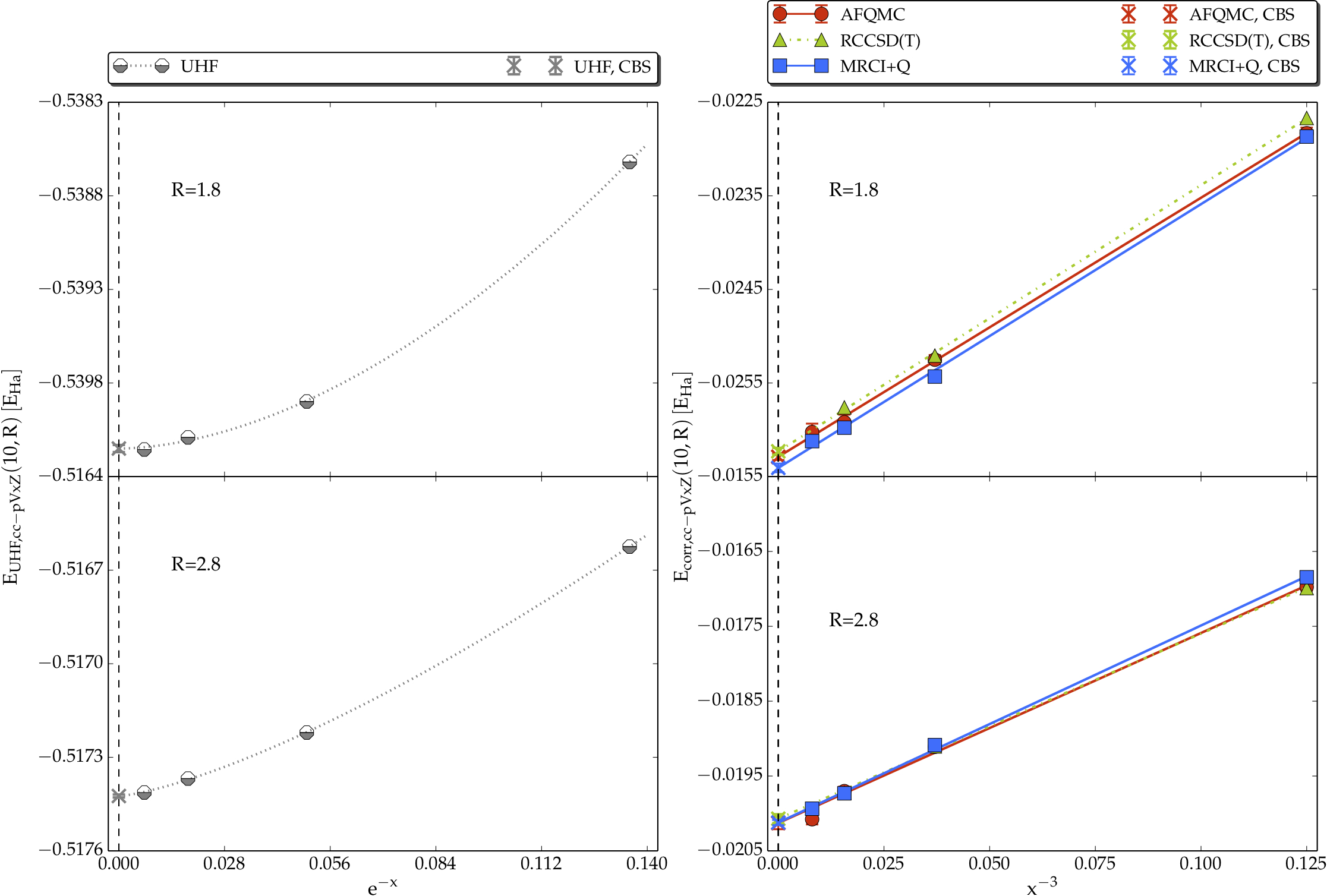}
    \caption{Extrapolation of the Hartree-Fock (left) and correlation energy (right) for chains of 10 hydrogen atoms \revision{separated by a distance of $R=1.8$ Bohr}, using cc-pVxZ Gaussian bases, and various methods from Ref.~\cite{motta2017towards}. \revision{Acronyms UHF, AFQMC, RCCSD(T) and MRCI+Q refer to unrestricted Hartree-Fock, auxiliary-field quantum Monte Carlo, restricted coupled cluster with singles and doubles and perturbative treatment of triples and multireference configuration interaction with Davidson correction respectively. The abbreviation CBS refers to the complete basis set limit ($x \to \infty$).}
    While the mean-field energy converges exponentially fast in the basis size (roughly given by $x^3$), the correlation energy converges as $x^{-3}$ (inversely proportional to the number of basis functions) due to the electron-electron cusp.}
    \label{fig:cusp}
\end{figure*}

In Gaussian bases, linear combinations of Gaussian functions (referred
to as simply Gaussian basis functions) are placed at the nuclear positions. As they are placed where
the ground-state electron density is highest, they give a compact representation
of the wavefunction for bound states, but the \revision{electron repulsion integral $u_{pqrs}$} is not sparse, with $O(N^4)$ second-quantized matrix elements.
In a quantum algorithm, this leads to high gate counts even for simple quantum primitives such as a Trotter step.

Plane-waves offer greater simplicity as the accuracy of the basis is controlled by a single parameter, the kinetic energy cutoff.
While the number of plane waves needed to reach a desired accuracy is larger than the number of Gaussian states, the Hamiltonian contains fewer terms ($O(N^3)$) due to momentum conservation. To reduce the number of required plane waves, it is essential to employ pseudopotentials
to remove the sharp nuclear cusp~\cite{martin2004electronic,marx2009ab}.
Furthermore, the asymptotic basis convergence of Gaussian and pseudopotential plane wave calculations is the same:
the feature governing the rate of convergence is the wavefunction discontinuity or \textit{electron-electron cusp} due to the singularity of the Coulomb interaction (see \revision{Figure~}\ref{fig:cusp}).
In classical simulations, so-called explicit correlation methods can be used to remove the slow convergence due to the singularity~\cite{klopper2006r12,kong2011explicitly}. How to use such techniques with quantum computers has yet to be explored.

The need to expose more sparsity in the Hamiltonian while retaining a reasonably compact wavefunctions is an active
area of research in both classical and quantum algorithms. Recent ideas have included new types of basis function 
that return to a more grid-like real-space basis~\cite{white2017sliced,white2019multisliced,babbush2018low,white2017hybrid,lin2012adaptive} where the Coulomb operator and thus Hamiltonian has only a quadratic number of terms,
as well as factorizations of the Coulomb operator itself~\cite{motta2018low,hohenstein2012tensor}.  The best choice of basis for a quantum simulation
remains very much an open question. 

\subsubsection{Fermion-to-qubit mappings}
\label{sec:f2q}

\begin{figure*}
\centering
\includegraphics[width=0.75\textwidth]{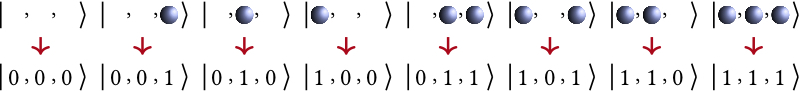}
\caption{Schematic of the Jordan-Wigner encoding for $3$ spin-orbitals. Fermions are represented by blue spheres.
Each Fock basis vector (from the vacuum state, left, to the completely filled state, right) is mapped onto a $3$-qubit 
state, with empty (filled) spin-orbitals corresponding to qubits in $0$ ($1$).
} \label{fig:jw}
\end{figure*}

\begin{figure}
    \centering
    \includegraphics[width=0.95\columnwidth]{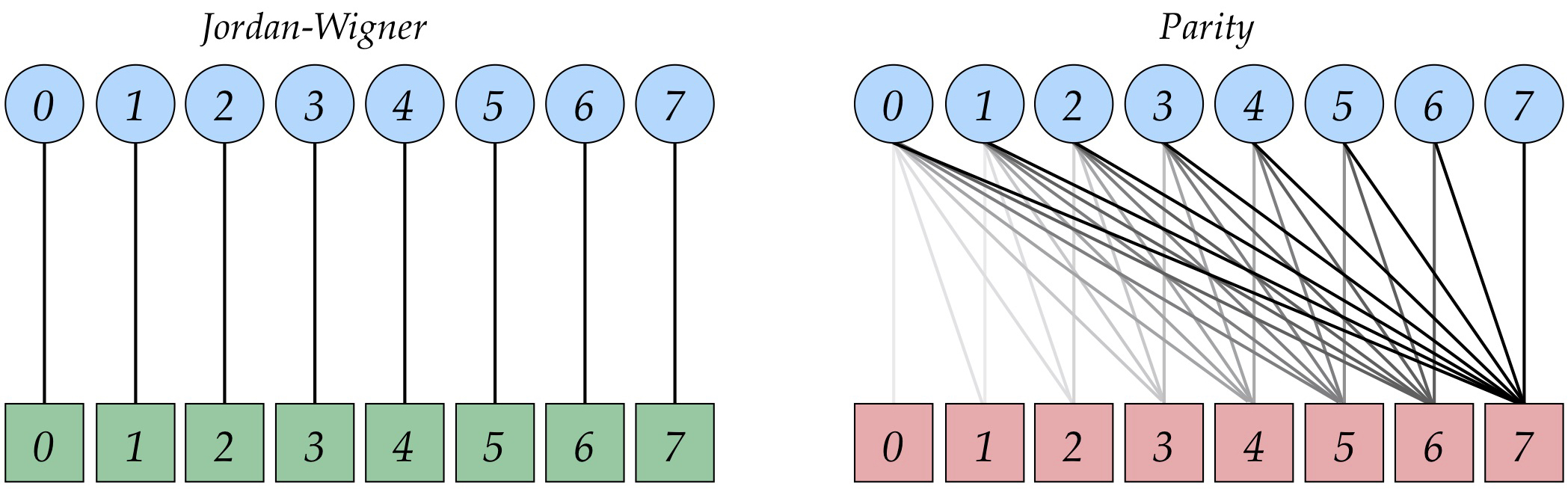}
    \caption{Difference between Jordan-Wigner and parity encoding. The former encodes occupation numbers $x_i$ on qubit states, the latter parities $p_i = \sum_{j<i} x_i \mbox{ mod } 2$. Other strategies, such as the binary-tree encoding, balance non-locality of occupation numbers and parities to achieve more efficient encodings.}
    \label{fig:parity}
\end{figure}

Since the basic units of a quantum computer are qubits rather than fermions, any quantum simulation algorithm of fermions 
(e.g. for electronic structure) employs a suitable encoding of fermionic degrees of freedom into qubits.
\revision{For example, the standard \textit{Jordan-Wigner} mapping \cite{1928jw} (sketched in Figure~\ref{fig:jw})} identifies each Fermi mode (orbital) 
with a qubit such that the empty and the occupied states  are mapped to the qubit basis states $|0\ra$ and $|1\ra$ respectively.
More generally, the Fock basis vector $|n_1,\ldots,n_N\ra$ is mapped to a qubit basis vector $|x_1\rangle \otimes \cdots \otimes |x_N\ra$,
where each bit $x_j$ stores a suitable partial sum (modulo two) of the occupation numbers $n_1,\ldots,n_N$. 
The Jordan-Wigner  mapping corresponds to $x_j=n_j$ for all $j$.
This is not quite satisfactory since single-mode creation/annihilation operators \revision{take the form $\hat{c}^\dagger_j \to \sigma^+_j Z_{j-1} \dots Z_1$ in the Jordan-Wigner representation. The product of $Z$ operators, referred to as a Jordan-Wigner string, is needed to reproduce canonical anticommutation relations between creation/annihilation operators, and thus to capture fermionic statistics. However, for increasing $j$, it becomes increasingly non-local in terms of qubits, with local fermionic operators acting in the worst case over a linearly increasing number of qubits with system size, leading
  to larger circuits and more measurements.}
On the other hand, updating the qubit state $x$ upon application of a single creation/annihilation operator 
requires a single bit flip (see \revision{Figure~}\ref{fig:parity}).
More efficient fermion-to-qubit mappings balance the cost of computing Jordan-Wigner strings and the bit-flip cost of updating the qubit state. 
For example, the \revision{Bravyi-Kitaev encoding~\cite{BK02}} maps any fermionic  single-mode operator
(e.g. $\hat{c}_p$ or $\hat{c}_p^\dag$) to a qubit operator acting non-trivially on roughly  $\log{N}$ or less qubits.
Generalizations of this encoding were studied in~\cite{havlivcek2017operator,seeley2012bravyi,tranter2015b}.
As a consequence, the second-quantized Hamiltonian Eq.~(\ref{Q2}) expressed in terms of qubits becomes a {linear combination of
Pauli terms} with weight at most $O(\log{N})$.
This is important in the context of  VQE-type quantum simulations (see \revision{Subsection}~\ref{sec:vqe}) since Pauli operators with an extensive
weight (of order $N$) cannot be measured reliably in the absence of error correction.

A natural question is whether the number of qubits required to express a Fermi system can be reduced by {exploiting symmetries} 
such as the particle number conservation or the point group symmetries of molecules.
For example, zero temperature simulations often target only one symmetry sector containing the ground state.
This motivates  the study of \textit{symmetry-adapted  fermion-to-qubit mappings}.
The goal here is to reduce the number of qubits required for the simulation 
without compromising the simple structure of the resulting qubit Hamiltonian (such as sparsity).
The simplest case of \revision{tapering (i.e. $Z_2$)} symmetries is now well understood
and the corresponding symmetry adapted mappings are routinely used
in experiments~\cite{bravyi2017tapering,o2016scalable,kandala2017hardware,hempel2018quantum}.
The $U(1)$ symmetry underlying  particle number conservation
was considered  in~\cite{moll2016optimizing,bravyi2017tapering,steudtner2018fermion}.

Dimension counting shows that
a system of $N$ Fermi modes with exactly $K$ particles can be mapped
to roughly $\log_2{N \choose K}$ qubits. However, it remains an open question whether
this mapping can be chosen such that the resulting qubit Hamiltonian
admits a sparse representation in some of the commonly used operator bases (such as the
basis of  Pauli operators) to enable applications in VQE.
Mappings adapted to point group symmetries
have been recently considered in~\cite{fischer2019symmetry}.
It is also of great interest to explore fermion-to-qubit mappings adapted to approximate and/or emergent symmetries.

Alternatively, one may artificially introduce symmetries either to the original Fermi system
or its encoded  qubit version with the goal of simplifying the resulting qubit Hamiltonian. 
This usually requires redundant degrees of freedom such as
auxiliary qubits or Fermi modes~\cite{BK02,ball2005fermions,verstraete2005mapping,setia2018superfast,steudtner2019quantum,jiang2019majorana}.
In the case of lattice fermionic Hamiltonians such as the 2D Fermi Hubbard model 
or more general models defined on bounded degree graphs,
such symmetry-adapted mappings produce a local qubit Hamiltonian 
composed of Pauli operators of constant weight independent of $N$. 

\subsubsection{Model and non-electronic problems}

While much of the above has focused on ab initio quantum chemistry and electronic structure in quantum simulations, the diverse
questions of chemistry and materials physics discussed in section \ref{sec:simulations} raise additional issues of representation.
For example, model Hamiltonians avoid the problem associated with a choice of basis by restricting the Hamiltonian to a predetermined simple form on a lattice. The
lattice structure permits specialized techniques, such as specific fermionic encodings. Developing specialized representations
for model problems is of particular importance in simulating condensed matter systems.

Other kinds of non-electronic simulations may involve different requirements on the basis than electronic problems. For example, in quantum
reactive scattering processes, there is little a priori information on the positions of the particles; instead various grid representations,
often in non-Cartesian coordinate systems, are used~\cite{colbert1992novel,pack1987quantum}. Alternatively, the particles of interest may be bosons
which engender new encoding considerations. Relatively little attention has been paid to these questions so far in quantum algorithms.

\subsection{Quantum algorithms for ground and excited states}
\label{sec:quantum_ground_state}

There are many approaches to obtaining ground states or excited states on a quantum computer. State preparation
procedures attempt to construct a circuit to prepare a state with as large as possible overlap with the desired eigenstate.
One set of such procedures, which we review in \revision{Subsection}~\ref{sec:gs_preparation} and which includes \textit{adiabatic state preparation} and \textit{quantum imaginary time evolution}, uses a prescribed evolution path.
An alternative strategy is based on \textit{variational methods} (often called variational quantum eigensolvers) and is reviewed in \revision{Subsection}~\ref{sec:vqe}.
Here, the preparation circuit itself is defined via the optimization of the energy with respect to parameters of the circuit.

Given some state with reasonably large overlap with the desired state, one can perform \textit{quantum phase estimation}, which simultaneously projects the state onto an eigenstate of the Hamiltonian and obtains an estimate for the energy of this eigenstate. The error depends inversely on the simulation time.
The probability of successfully projecting onto the desired state is given by the square overlap of the input state and the desired state, and it is thus necessary to use some other method (such as the state preparation procedures above) to prepare an input state with
sufficient overlap with the desired state.

The various approaches
come with different {strengths and weaknesses}. While phase estimation allows the deviation of the final state from an exact eigenstate (although not necessarily the desired eigenstate) to be systematically reduced, it can require deep circuits with
many controlled gates that are challenging for devices with limited coherence and without error correction.
Variational methods or quantum imaginary time evolution replace such circuits by a large number
of potentially shorter simulations, which is expected to be easier to implement on near-term machines. Finally, one can consider hybrid approaches~\cite{wang2019accelerated}.
However, if one does not measure the energy by phase estimation, but instead 
by expressing the Hamiltonian as a sum of multi-qubit Pauli operators and measuring the terms individually,  the state preparation and measurements
must be repeated many times, with the error converging only as the square root of the number of repetitions.
Variational methods are also limited by the variational form and ability to solve the associated optimization problem, which may by itself represent a difficult classical optimization. Finally, adiabatic state preparation and quantum imaginary time evolution become inefficient for certain Hamiltonians.

State preparation methods are first discussed in \revision{Subsection}~\ref{sec:gs_preparation} and \revision{Subsection}~\ref{sec:vqe} (variational state
preparation is discussed separately due to the large number of different types of ansatz).
 Considerations for excited states are
discussed in \revision{Subsection}~\ref{sec:excited}. Phase estimation is reviewed in \revision{Subsection}~\ref{sec:qpe}. 

\subsection{Preparing ground states along a prescribed path}
\label{sec:gs_preparation}

\subsubsection{Adiabatic state preparation}

\revision{One general route to prepare the ground-state of a physical system on a quantum device is through adiabatic state preparation. This relies on the well-known \textit{adiabatic theorem}~\cite{Born1928,Messiah1962}, which states that a system that starts in the ground-state of
some Hamiltonian at time $t=0$, will stay in the instantaneous ground-state of the time-dependent Hamiltonian $\hat{H}(t)$ under evolution
by $\hat{H}(t)$, so long as the evolution is sufficiently slow and the spectrum of the Hamiltonian remains gapped.} To make use of this, one chooses a Hamiltonian path $\hat{H}(\lambda)$, $0 \leq \lambda \leq 1$, such that the ground state of $\hat{H}(0)$ is easily prepared, while $\hat{H}(1)$ is the Hamiltonian whose ground state one wants to obtain. The system is then evolved under the time-dependent Schr\"{o}dinger equation,
\begin{equation}
i \, \frac{d}{dt} |\Phi_t \rangle = \hat{H}(t/T) |\Phi_t \rangle
\;,\;
0\leq t\leq T
\;
\end{equation}
In the limit $T \to \infty$ and if the spectrum of $\hat{H}_{t/T}$ is gapped for all $t$, the final state $|\Phi_T\rangle$ is the exact ground state of $\hat{H}$. 

\revision{In particular, it is known \cite{Messiah1962,veis2014adiabatic} that the time needed to ensure convergence to the ground state is $T > \epsilon/ g^{2}_{min}$. Here, 
$g_{min} = \min_{\lambda} E_1(\lambda) - E_0(\lambda)$ is the minimum gap along the adiabatic path, expressed 
in terms of the lowest two eigenvalues of $H(\lambda)$, $E_k(\lambda) = \langle \psi_k(\lambda) | \hat{H}(\lambda) | \psi_k(\lambda) \rangle$, and $\epsilon = \min_\lambda \left| \langle \psi_1(\lambda) | H^\prime(\lambda) | \psi_0(\lambda)  \rangle \right|$.
For Hamiltonians where the quantity $\epsilon/ g^{2}_{min}$ remains finite, or grows polynomially with system size, $T$ is roughly bounded by a constant or a polynomial function of system size, enabling simulation by adiabatic state preparation.
In determining the behavior of $\epsilon/ g^{2}_{min}$, an important role is played of course by the Hamiltonian gap. Hamiltonians that remain gapped,
or whose gap closes polynomially with system size, can typically be simulated in the adiabatic limit at polynomial cost in system size.}

Away from the adiabatic limit $T \to \infty$, corrections arise that depend on the instantaneous gap of $\hat{H}(\lambda)$ and the total time $T$. 
Furthermore, if degeneracies occur along the path, a different time-dependent Hamiltonian path (although with the same endpoints) must be chosen.

\begin{figure}
    \centering
    \includegraphics[width=0.95\columnwidth]{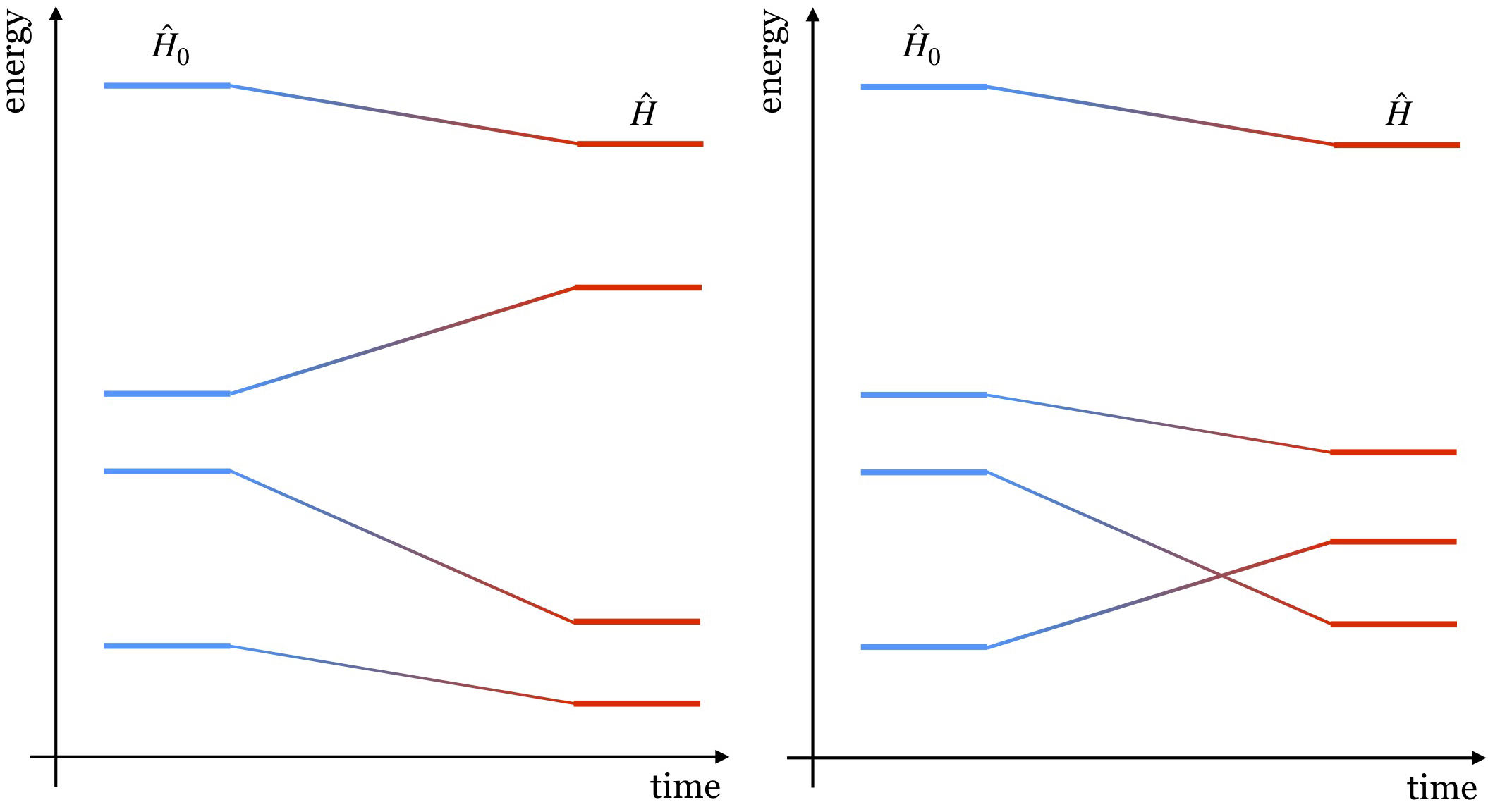}
    \caption{Adiabatic state preparation. Starting from an eigenstate of a simple Hamiltonian $\hat{H}_0$ and slowly switching on the
    interaction $\hat{H} - \hat{H}_0$ leads to an eigenstate of $\hat{H}$ (left).
    Along paths with degeneracies, adiabatic state preparation can lead to the wrong eigenstate.}
    \label{fig:adiabatic}
\end{figure}

\revision{Thus, while} this approach is {very general}, its practical {applicability is limited} by the requirement of having to take the limit of large $T$, and
to choose a path without degeneracies (see \revision{Figure}~\ref{fig:adiabatic}). The limit of large $T$ may require deep circuits that may not be practical in near-term quantum machines.
Analyzing the errors (and optimizing the path, for example by choosing an improved $f(s)$) is also challenging due to the dependence on the unknown spectrum of $\hat{H}_{t/T}$. Some of these questions have been studied in the more general context of adiabatic quantum computation~\cite{albash2018adiabatic}.
However, while this is one of the first state preparation methods discussed for chemical systems~\cite{aspuru2005simulated}, more heuristic work in this area for problems of interest to chemistry and physics is needed~\cite{veis2014adiabatic}.

\subsubsection{Quantum imaginary-time evolution}

In classical simulations, one popular approach to prepare (nearly exact) 
ground-states is imaginary-time evolution, which 
expresses the ground-state as the long-time limit of the imaginary-time Schr\"{o}dinger equation,
\begin{equation}
| \Psi \rangle = \lim_{\beta \to \infty} \frac{ e^{- \beta \hat{H} } | \Phi \rangle}{ \|  e^{- \beta \hat{H} } \Phi \| } \quad.
\end{equation}
Imaginary time-evolution underlies the family of projector quantum Monte Carlo methods in classical algorithms~\cite{martin2016interacting}. 

To perform imaginary time evolution on a quantum computer, it is necessary to implement the (repeated) action of
the short-imaginary-time propagator $e^{-\Delta \tau \hat{H}}$ on a state. Given a Hamiltonian that
can be decomposed into geometrically local terms, $\hat{H} = \sum_m \hat{h}_m$, and
a state $|\Psi\rangle$ with finite correlation length $C$, the action of $e^{-\Delta \tau \hat{h}_i}$
can be generated by a unitary $\hat{U} = e^{i\hat{A}}$ acting on $O(C)$ qubits surrounding those acted on by $\hat{h}_i$, i.e.
\begin{equation}
\frac{ e^{- \Delta\tau \hat{h}_i} | \Psi \rangle }{ \| e^{- \Delta\tau \hat{h}_i} \Psi \| } = \hat{U} | \Psi \rangle = e^{i \hat{A}} | \Psi \rangle \quad,
\end{equation}
where the coefficients of the Pauli strings in $\hat{A}$ can be determined from local measurements
of the qubits around $\hat{h}_i$. This is the idea behind the
quantum imaginary time evolution (QITE) algorithm~\cite{Motta2019}.
Like adiabatic state preparation, quantum imaginary time evolution can in principle prepare exact states without the
need for  variational optimization. 
Also, the total length of imaginary time propagation to achieve a given error is determined  by the spectrum of $\hat{H}$
and the initial overlap, rather than by the spectrum of $\hat{H}_{t/T}$ along the adiabatic path.
However, the method  becomes inefficient in terms of the number of measurements and complexity of the operator $\hat{A}$
if the domain $C$ grows to be large along the imaginary time evolution path. In these cases, QITE can be used
as a heuristic for approximate ground-state preparation, analogous to using
adiabatic state preparation for fixed evolution time. While initial estimates in a limited set of problems show QITE to be resource efficient
compared to variational methods due to the lack of an optimization loop~\cite{Motta2019},  a better numerical understanding
of its performance and cost across different problems, as well as the accuracy of inexact QITE in different settings, remains to be developed.

\subsection{Variational state preparation and variational quantum eigensolver}

\label{sec:vqe}
A class of state preparation methods that have been argued
to be particularly amenable to near-term machines is variational state preparation~\cite{farhi2014quantum,Peruzzo2013,mcclean2016theory,romero2018strategies}. Here, similar to classical variational approaches, one chooses a class of ansatz states for the ground state of the Hamiltonian of interest. Generally speaking, such an ansatz consists of some initial state and a unitary circuit parametrized by some set of classical {variational parameters}. Applying this circuit to the initial state yields a guess for the ground state, whose energy is then evaluated. This yields an upper bound to the true ground state energy. One then varies the variational parameters to lower the energy of the ansatz state.

In choosing the class of ansatz states, one pursues several goals: on the one hand, it is crucial that the class contains an accurate approximation to the true ground state of the system. On the other hand, one desires a class of circuits that are easily executed on the available quantum computer, i.e. for a given set of available gates, connectivity of the qubits, etc. Finally, it is important for the classical optimization over the variational parameters to be well-behaved, \revision{so as to be able to find low-energy minima \cite{mcclean2018barren,akshay2020reachability}}. While we cannot list all possible ansatz states below, we provide a representative sample.

\subsubsection{Unitary coupled cluster}

\begin{figure}
    \centering
    \includegraphics[width=0.75\columnwidth]{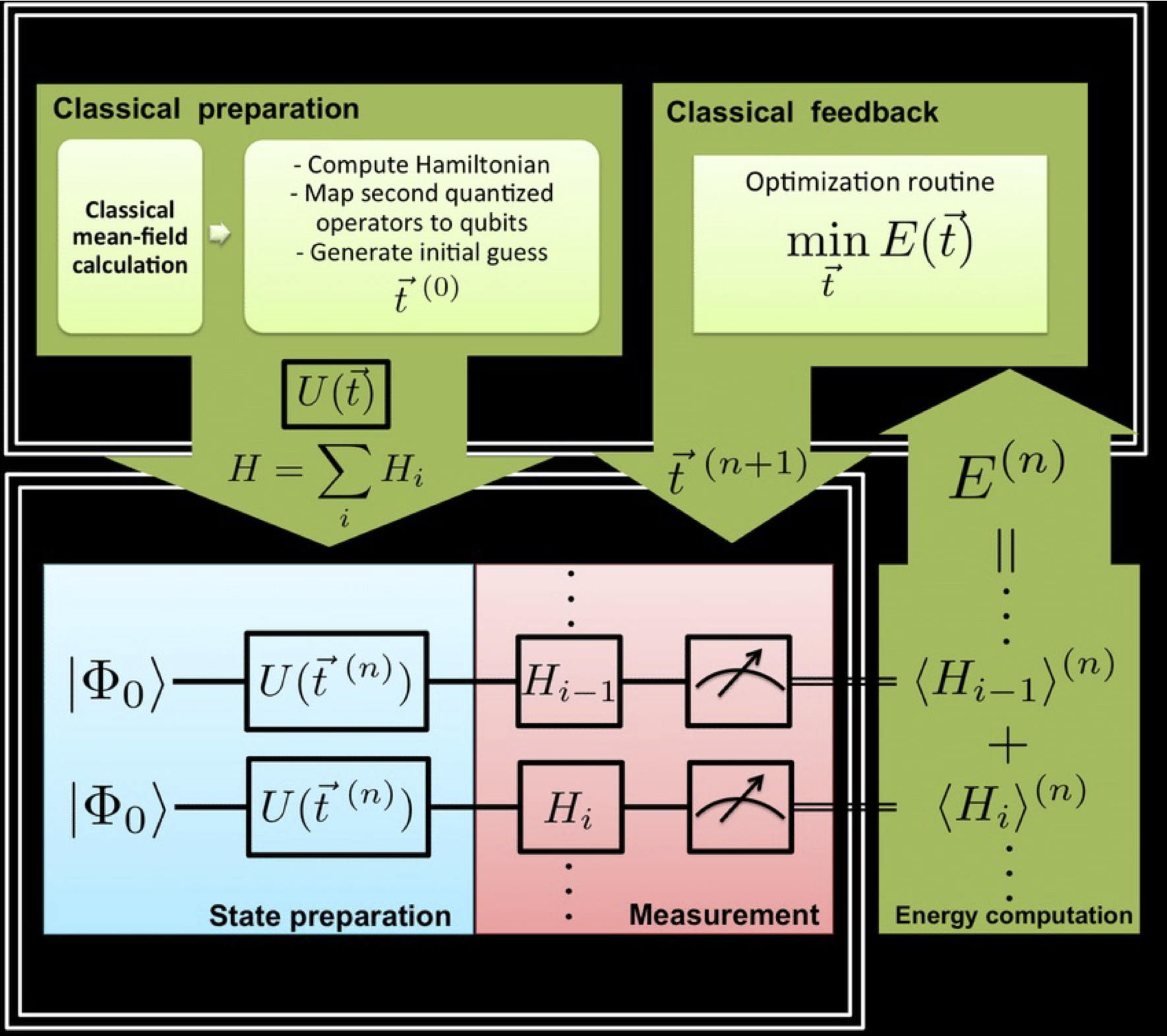}
    \caption{Workflow of the variational quantum eigensolver algorithm. The classical optimization routine adds expectation values of the Hamiltonian
    Pauli terms to calculate the energy and estimates new values for the unitary parameters. The process is repeated until convergence. From \cite{romero2018strategies}.}
    \label{fig:vqe}
\end{figure}

An early example of a particular class of ansatz states that has been suggested for applications in quantum chemistry is the 
unitary coupled-cluster (uCC) ansatz~\cite{bartlett1989alternative,Peruzzo2013,Moll_2018,romero2018strategies,albash2018adiabatic}
\begin{equation}
\begin{split}
| \Psi_{uCC} \rangle &= e^{ \hat{T} - \hat{T}^\dag } | \Psi_{HF} \rangle
\;,\; \\
\hat{T} &= \sum_{k=1}^d \sum_{ \substack{i_1 \dots i_k \\ a_1 \dots a_k}} t^{a_1 \dots a_k}_{i_1 \dots i_k}
\hat{c}_{a_k}^\dagger \dots \hat{c}_{a_1}^\dagger \hat{c}_{i_k} \dots \hat{c}_{i_1}
\;.
\end{split}
\end{equation}
Here, $d$ denotes the maximum order of excitations in the uCC wavefunction
(for example $d=1$, $2$, $3$ for singles, doubles and triples respectively),
$\hat{c}_{a_k}^\dagger \dots \hat{c}_{a_1}^\dagger$ ($\hat{c}_{i_k} \dots \hat{c}_{i_1}$)
are creation (destruction) operators relative to orbitals unoccupied (occupied)
in the Hartree-Fock state, and $t$ is a rank-$2k$ tensor, antisymmetric in the
$a_k \dots a_1$ and $i_k \dots i_1$ indices. This choice of ansatz is motivated
by the {success of mean-field theory}, which suggests that the density
of excitations in the true wavefunction should be small relative to the mean-field state. Standard
coupled cluster theory -- written as $e^{\hat{T}}|\Psi_{HF}\rangle$ --
is widely used in classical quantum chemistry
but is challenging to implement on a quantum device, whereas the reverse is true for the unitary variant.
Understanding the theoretical and numerical differences between standard and unitary coupled cluster
is an active area of research~\cite{Cooper2010,harsha2018difference}.
The variational quantum eigensolver algorithm applied to the unitary coupled-cluster Ansatz is depicted in \revision{Figure}~\ref{fig:vqe}.

\subsubsection{Hardware-efficient ansatz}

\begin{figure*}
    \centering
    \includegraphics[width=0.6\textwidth]{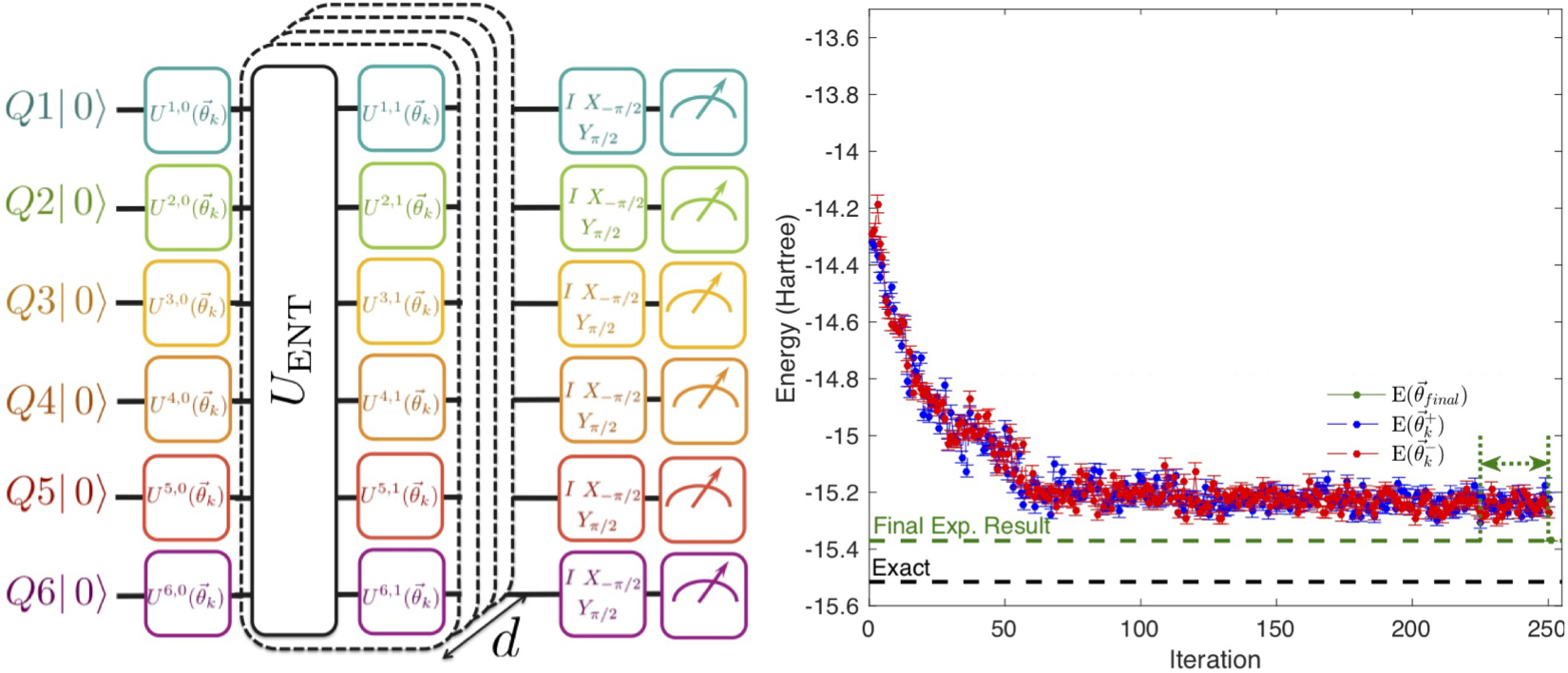}
    \caption{Left: Hardware-efficient quantum circuit for trial state preparation and energy estimation, shown here for 6 qubits. The circuit is composed of a sequence of interleaved single-qubit rotations, and entangling unitary operations UENT that entangle all the qubits in the circuit. A final set of post-rotations prior to qubit readout are used to measure the expectation values of the terms in the qubit Hamiltonian, and estimate the energy of the trial state. Right: energy minimization for the six-qubit Hamiltonian describing BeH$_2$. Adapted from Ref.~\cite{kandala2017hardware}.}
    \label{fig:hardware_efficient}
\end{figure*}

The unitary coupled cluster ansatz involves non-local gate operations and is expensive to implement on near-term
devices with limited qubit connectivity. 
An alternative variational approach, pursued e.g. in Ref.~\cite{kandala2017hardware} and termed ``hardware-efficient'' there, is to tailor the ansatz specifically to the underlying hardware characteristics.
The circuits considered in Ref.~\cite{kandala2017hardware}, sketched in \revision{Figure}~\ref{fig:hardware_efficient}, consist of alternating layers of arbitrary single-qubit gates and an entangling gate that relies on the intrinsic drift Hamiltonian of the system. While this drift Hamiltonian and thus the entangling gate is not known precisely, for the variational approach it is sufficient to know that the gate is reproducible. The variational parameters are only the rotations in the layer of single-qubit gates. While it is not guaranteed that such an ansatz contains a good approximation to the state of interest, it is an example
of an adaption of a method to NISQ devices~\cite{kandala2017hardware}.
An application to BeH$_2$ is seen in \revision{Figure}~\ref{fig:hardware_efficient}.

\subsubsection{Adapt-VQE ansatz}

In the adapt-VQE scheme, a collection of operators $\hat{A}_i$ (operator pool) is chosen in advance, and the ground state is approximated by 
\begin{equation}
| \Psi_{\mbox{adapt-VQE}} \rangle = e^{ \theta_n \hat{A}_n} \dots e^{ \theta_1 \hat{A}_1} | \Psi_{\mbox{HF}} \rangle \quad,
\end{equation}
Given a current parameter configuration $\theta_1$ $\dots$ $\theta_n$, the commutator of the Hamiltonian with
each operator in the pool is measured to obtain the gradient of the energy
\begin{equation}
E = \langle \Psi_{\mbox{adapt-VQE}} | \hat{H} | \Psi_{\mbox{adapt-VQE}} \rangle
\end{equation} 
with respect to the parameters $\theta$. Repeating this multiple times and averaging over the obtained samples gives the gradient of the expectation value of the Hamiltonian with respect to the coefficient of each operator. The ansatz is improved by adding the operator $\hat{A}_i$ with the largest gradient to the left end of the
ansatz with a new variational parameter, thereby increasing $n$.
The operation is repeated until convergence of the energy \cite{Grimsley2019}.
Numerical simulations, for example for short hydrogen chains, show that adapt-VQE can improve over the unitary coupled cluster approach in terms of
the accuracy reached for a given circuit depth.

\subsubsection{Tensor networks}

\begin{figure*}
    \centering
    \includegraphics[width=0.6\textwidth]{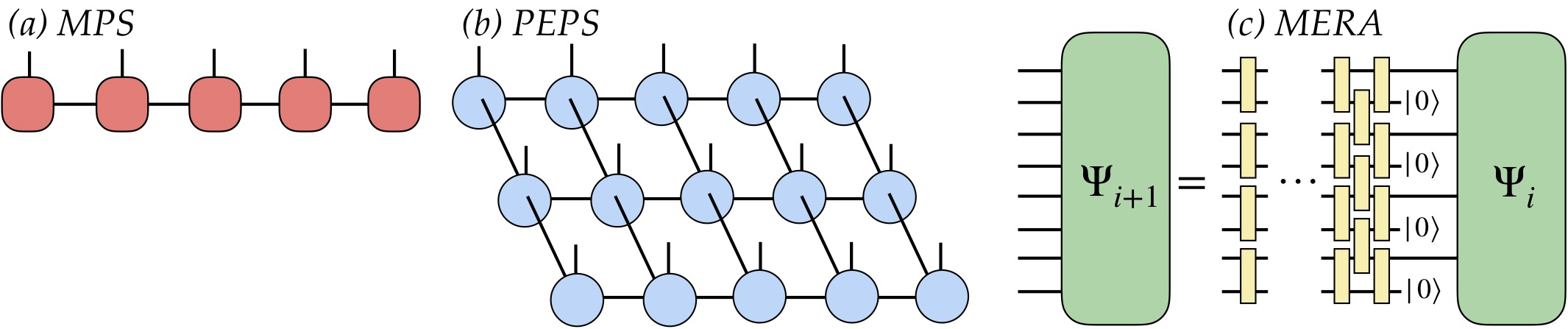}
    \caption{Graphical representation of (a) matrix product states, (b) projected-entangled pair states. The boxes and circles represent
    tensors of numbers.  (c) Construction of deep multi-scale entanglement renormalization ansatz (deep MERA). The figure
    shows an isometry (one component of the MERA ansatz) being constructed with exponential dimension via quantum circuits.
    Adapted from Ref.~\cite{orus2014practical} and Ref.~\cite{kim2017robust}.}
    \label{fig:tns}
\end{figure*}

Tensor networks are {a class of variational states} which construct the global wavefunction amplitude from tensors associated with local degrees
of freedom. They specify a class of quantum states that can be represented by an amount of classical information proportional to the
system size. There are two main families of tensor networks: those based on \textit{matrix product states} and
\textit{tree tensor network states} (MPS, TTNS)~\cite{white1992density,MPSfannes}
(also known in the numerical multi-linear algebra community as the tensor-train decomposition \cite{oseledets2009breaking,oseledets2011tensor}
and the tree-structured hierarchical Tucker representation \cite{hackbusch2009new,grasedyck2010hierarchical})
and their higher dimensional analogs, \textit{projected entangled pair states} (PEPS)~\cite{PEPS}; and those based on the \textit{multi-scale entanglement renormalization ansatz} (MERA)~\cite{MERAvidal}. Because of their success in representing low-energy states in classical simulations, they are a natural class of variational wavefunctions to try to prepare in a quantum algorithm for low-energy states. These tensor networks are schematically depicted in \revision{Figure}~\ref{fig:tns}.

There are many analogies between tensor network algorithms and quantum circuits. This analogy can be exploited
to develop an efficient preparation mechanism for these states on a quantum computer. By recognizing that the tensors in an MPS or a MERA
can be associated with a block of unitaries (with the bonds between tensors playing the role of circuit lines in a quantum circuit) it is possible
to prepare an MPS or MERA state on a quantum computer~\cite{schon2007sequential,huggins2018towards,liu2019variational}. Because the dimensions of the associated tensor grow exponentially with the depth
of the quantum circuit associated with it, it is possible to prepare certain {tensor networks with large bond dimension} on a quantum computer
that {presumably cannot be efficiently simulated classically}; an example of this is the so-called \textit{deep MERA}~\cite{kim2017robust}.

There are many {open questions} in the area of tensor networks and quantum computing. For example, preparing PEPS on a quantum computer appears
to be much less straightforward than preparing a matrix product state or a MERA~\cite{schwarz2012preparing}. Similarly, although ``deep'' tensor network states can only be efficiently simulated on a quantum computer, their additional representational power over
classically efficient tensor networks for problems of physical or chemical interest is poorly understood.

\subsubsection{Other considerations}

Besides the choice of ansatz state, the computational challenges of variational methods and VQE are twofold: 
\begin{itemize}
\item Potentially, a very {large number of measurements} must be performed to accurately estimate the energy. Indeed, the scaling is quadratically worse than when using quantum phase \revision{estimation, in a simple implementation. Recently, however, considerable effort has been put in devising schemes to reduce the number of measurements required to estimate the energy, without sacrificing accuracy \cite{izmaylov2019revising,torlai2020precise,bonet2019nearly,izmaylov2019unitary,zhao2020measurement}.} We will discuss this point further in \revision{Section}~\ref{sec:readout}.
\item The {optimization} of the variational parameters may be very difficult, in particular if the energy exhibits a very non-trivial dependence on classical parameters with many local minima, and if {gradient information} is not easily available. For some discussion of optimization algorithms in this context, see Refs.~\cite{mcclean2015theory,yang2017,mcclean2018barren,akshay2020reachability,stokes2020quantum}.
\end{itemize}

Some key advantages of the VQE approach are that it can often be carried out with a large number of {independent, short quantum simulations}. This is more suitable to NISQ machines than the long coherent circuits required for approaches based on quantum phase estimation, which has been demonstrated in several experiments~\cite{kandala2017hardware,o2016scalable}. Furthermore, the approach is more {resilient against certain types of errors}. For example, as mentioned already above, it is generally not necessary to know exactly what circuit is executed for some variational parameters as long as it is reproducible; therefore, systematic coherent tuning errors of the qubits (for example systematic deviations between the desired and the actually applied single-qubit rotations) do not adversely affect the results. 
In addition to studying the robustness of VQE against errors, it has become a very active field to develop techniques that {mitigate such physical errors}. Such  approaches promise to reduce the impact of errors on near-term machines before error correction becomes available. For work in this direction, see Refs.~\cite{mcclean2015theory,mcclean2017,LB2016,temme2016error,endo2018,bonet2018,mcardle2019}.

\subsection{Excited states}

\label{sec:excited}
While much of the above discussion of state preparation and variational algorithms has focused on ground-states,
most of the same methods can also be used with minor extensions for excited states. For example, adiabatic state preparation
can be used to prepare an excited state, so long as it is connected to the initial state without a vanishing gap.

In the area of variational methods, it is often useful to choose the excited state ansatz
to be related to that of the ground-state, since at low-energies much of the physics is the same.
This is widely used in classical simulations and essentially the same ideas have been ported to the quantum algorithm setting.
For example, in the \textit{quantum subspace expansion} (QSE)~\cite{colless2018computation}, the excited state is made via the ansatz $|\Psi'\rangle = \sum_\alpha c_\alpha \hat{E}_\alpha |\Psi\rangle$, where $\{ E_\alpha\}$ is a set of ``excitation'' operators and $|\Psi\rangle$ is the ground-state constructed in VQE. In QSE, one needs
to measure all the subspace matrix elements $\langle \Psi| \hat{E}_\alpha^\dag \hat{E}_\beta |\Psi\rangle$, $\langle \Psi| \hat{E}_\alpha^\dag \hat{H} \hat{E}_\beta |\Psi\rangle$, thus the number of measurements
grows quadratically with the subspace. 
In the \textit{quantum Lanczos} method, the QITE algorithm is used to construct the subspace $\{ e^{-\lambda \hat{H}} |\Psi\rangle,
e^{-2\lambda \hat{H}}|\Psi\rangle, \ldots \}$ and the special structure of this space means that all subspace matrix elements
can be constructed with a number of measurements that grows only linearly with the
size of the subspace~\cite{Motta2019}. Alternatively, one can fix the coefficients $c_\alpha$ and reoptimize the quantum circuit in the variational method; this is the basis of the multi-state VQE method; other similar ideas have also been proposed. Connections between quantum subspaces and error correction have been explored in  \cite{QSE_ec}.

The above methods compute total energies of excited states, which have to be subtracted from the ground state energy
to give the excitation energies of the system. 
A method to directly access excitation energies is desirable.
One route to achieve this goal is provided by the \textit{equation-of-motion} (EOM) approach, also widely used in classical simulations \cite{Rowe} and recently extended to quantum computing \cite{Ganzhorn}.
In the EOM approach, excitation energies are obtained as 
\begin{equation}
\Delta E_n = \frac{ \langle  \Psi | [\hat{O}_n,\hat{H},\hat{O}_n^\dag] | \Psi \rangle }{ \langle  \Psi | [ \hat{O}_n ,\hat{O}_n^\dag] | \Psi \rangle }
\end{equation} 
where \revision{$[\hat{O}_n,\hat{H},\hat{O}_n^\dag]=([[\hat{O}_n,H_n],O_n^\dag] + [\hat{O}_n,[H_n,O_n^\dag]])/2$ is a  double commutator},
$\Psi$ is an approximation to the ground state (such as the VQE ansatz) and $\hat{O}_n$ is an excitation operator expanded on a suitable basis.
The variational problem of finding the stationary points of $\Delta E_n$ leads to a generalized eigenvalue equation,
the solutions of which are the excited-state energies.

\subsection{Phase estimation}
\label{sec:qpe}

\begin{figure}
    \centering
    \includegraphics[width=0.75\columnwidth]{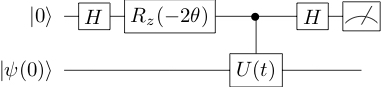}
    \caption{Quantum circuit for iterative QPE. The first two single-qubit gates bring the ancilla to the state $\frac{1}{\sqrt{2}} \, \left( |0 \rangle + e^{i\theta} |1\rangle \right)$. The controlled-$U(t)$ gate, with $U(t) = e^{-i t \hat{H}}$, brings the register into the state \revision{Eq.~}\eqref{eq:peastate}. The last Hadamard gate is needed to measure in the $X$ basis.}
    \label{fig:ipea}
\end{figure}

Quantum Phase Estimation (QPE) is a crucial step in many quantum algorithms.
In the context of quantum simulation, QPE enables {high-precision measurements of} 
the ground and excited {energy levels}.  This is achieved by preparing a trial initial state
$|\psi(0)\rangle$ 
that has a non-negligible overlap with the relevant eigenvector 
of the target Hamiltonian $\hat{H}$
and 
applying a quantum circuit that 
creates a superposition of 
time evolved 
states $|\psi(t)\rangle = e^{-i\hat{H}t}|\psi(0)\rangle$ over a suitable
range of the evolution times $t$.
In the simplest case, known as the {iterative QPE}~\cite{kitaev2002classical,aspuru2005simulated} and sketched in \revision{Figure}~\ref{fig:ipea}, the final state  is a superposition of the initial state itself
and a single time-evolved state,
\begin{equation}
\label{eq:peastate}
|\psi_\theta(t) \rangle =  \frac1{\sqrt{2}} \left( |0\rangle \otimes |\psi(0)\rangle + 
e^{i\theta}
|1\rangle \otimes |\psi(t)\rangle \right).
\end{equation}
Here one ancillary control qubit has been added that determines whether
each gate in the quantum circuit realizing 
 $e^{-i\hat{H}t}$ is turned on (control is $1$) or off (control is $0$).
The extra phase shift $\theta$ coordinates interference between the two
computational branches such that useful information can be read out with high confidence.
Finally, the control qubit is measured in the so-called  $X$-basis,
$|\pm\rangle =(|0\rangle \pm |1\rangle)/\sqrt{2}$ 
and the measurement outcome
$b\in \{+1,-1\}$ is recorded. 

The iterative QPE works by performing many runs of the above subroutine
with a suitable choice of parameters $t$, $\theta$ in each run and performing
a classical post-processing of the observed measurement outcomes.
The ancillary control qubit stays alive only over the duration of a single run since its
state is destroyed by the measurement. However, the remaining
qubits that comprise the simulated system stay alive over 
the entire duration of the QPE algorithm. More specifically, let $E_\alpha$ and $|\psi_\alpha\rangle$ be the eigenvalues and
eigenvectors of $\hat{H}$ such that  $\hat{H}=\sum_\alpha E_\alpha |\psi_\alpha\rangle \langle \psi_\alpha|$.
The trial state can be expanded in the eigenbasis of $\hat{H}$ as 
$|\psi(0)\rangle = \sum_\alpha c_\alpha |\psi_\alpha\rangle$.
Then the joint probability distribution describing
measurement outcomes $b_1,\ldots,b_N=\pm 1$ observed in $N$ 
runs of QPE has the form 
\begin{equation}
\mathrm{Pr}(b_1,\ldots,b_N) =  \sum_\alpha |c_\alpha|^2 \prod_{i=1}^N 
 \frac12\left( 1 +b_i  \cos{(\theta_i - E_\alpha t_i)}\right).
\end{equation}
Here $\theta_i$ and $t_i$ are the phase shift and the evolution time used in the $i$-th run
and $b_i=\pm 1$ is the observed measurement outcome.
This has the same effect as picking an eigenvector  of $\hat{H}$ at random 
with the probability $|c_\alpha|^2$ and then running QPE on the initial
state $|\psi(0)\rangle = |\psi_\alpha\rangle$.
Accordingly, QPE aims at estimating a {\em random}
eigenvalue $E_\alpha$ sampled from the probability distribution $|c_\alpha|^2$. 
We note that  a suitably formalized version of this problem 
with the trial state
$|00\ldots0\rangle$ and a local Hamiltonian $\hat{H}$ composed of few-qubit
interactions is known to be $\mathsf{BQP}$-complete~\cite{wocjan2006several}. 
In that sense, QPE captures the full computational power of quantum
computers and any quantum algorithm can be expressed as a special case
of QPE.
A common application of QPE is the task of estimating the smallest eigenvalue $E_0=\min_\alpha E_\alpha$.
This requires a trial state $|\psi(0)\rangle$ that has a non-negligible overlap
with the true ground state of $\hat{H}$ to ensure that the minimum of a few randomly sampled eigenvalues
$E_\alpha$ coincides with $E_0$.
For example, in the context of molecular
simulations,  $|\psi(0)\rangle$ is often chosen as the Hartree-Fock 
approximation to the ground state. Such a state is easy to prepare as it can be chosen to correspond
to a standard basis vector, see \revision{Subsection}~\ref{sec:f2q}.

The problem of {obtaining a good estimate of the eigenvalue} $E_\alpha$
based on the measured outcomes $b_1,\ldots,b_N$ is an active {research area},
see~\cite{svore2013faster,wiebe2016efficient,kimmel2015robust,o2019quantum}.
Assuming that the trial state has a constant overlap with the
ground state of $\hat{H}$, the smallest eigenvalue $E_0$
can be estimated using QPE
within a given error $\epsilon$ using $N\sim \log{(1/\epsilon)}$ runs
such that each run evolves the system over time at most $O(1/\epsilon)$.
As discussed in \revision{Subsection}~\ref{sec:quantum_time_evolution},
the evolution  operator $e^{-i\hat{H}t}$ (as well as its controlled version)
can be approximated by a quantum circuit of size scaling almost linearly in $t$
(neglecting logarithmic corrections). Thus QPE can achieve
an approximation error $\epsilon$ at the computational cost roughly $1/\epsilon$
even if the trial state has only a modest overlap with the 
ground state. This should be contrasted with VQE algorithms that have  cost at least
$1/\epsilon^2$ due to sampling errors 
and where the trial state must be a very good approximation to the
true ground state, see \revision{Subsection}~\ref{sec:vqe}.
On the other hand, QPE is much more demanding in terms
of the required {circuit depth and the gate fidelity}. It is expected that
{quantum error correction will be required} to implement QPE in a useful way (e.g. to outperform VQE in ground-state determination).

QPE  also has a {single-run (non-iterative) version}
where the time evolution of the simulated system  is controlled by a multi-qubit register
and the $X$-basis measurement of the control qubit is replaced by the 
Fourier basis measurement~\cite{abrams1999quantum}.
The iterative version of QPE has the clear advantage of requiring
fewer qubits. It also trades quantum operations required to realize the Fourier basis measurement
for classical postprocessing, thereby reducing the overall quantum resource cost of 
the  simulation.

Since {QPE is used ubiquitously} in a variety of quantum applications,
 it is crucial to {optimize its performance}. Below we list some open problems
 that are being actively investigated; see Ref.~\cite{o2019quantum} for a recent review.
 \begin{itemize}
\item Given limitations of near-term quantum devices,
 of particular interest are {tradeoffs between the depth} of the QPE circuit and its {spectral-resolution power}
 as well as its {sensitivity to noise}.
 It was shown~\cite{o2019quantum} 
 that that the computational cost of QPE interpolates between $1/\epsilon$ and $1/\epsilon^2$
 as the depth (measured by the number of runs $N$ per iteration of the algorithm) is reduced
 from $O(1/\epsilon)$ to $O(1)$. A particular version of QPE with a tunable depth
 that interpolates between the standard iterative QPE and VQE
 was proposed in~\cite{benedetti2019parameterized}.
\item Several methods have been proposed for {mitigating experimental errors} for VQE-type
 simulations~\cite{temme2016error,bonet2018,mcardle2019}.
 Such methods enable reliable estimation of expected values of observables
 on a given trial state without introducing any overhead in terms of extra qubits or quantum gates.
 Generalizing such error mitigation methods to QPE is a challenging open problem since QPE
 performs a non-trivial postprocessing of the measurement outcomes
 that goes beyond computing mean values. 
\item Classical {post-processing} methods that enable simultaneous estimation
 of multiple eigenvalues are highly desirable~\cite{o2019quantum}.
\item Finally, a natural question is whether the time evolution operator $e^{-i\hat{H}t}$ in QPE
 can be replaced by some other functions of $\hat{H}$ that are {easier to implement}~\cite{poulin2018quantum,berry2018improved}.
\end{itemize}

\subsection{Quantum algorithms for time evolution}

\label{sec:quantum_time_evolution}

\subsubsection{Hamiltonian simulation problem}

It was recognized early on~\cite{feynman1982simulating,meyer1996quantum,wiesner1996simulations,Lloyd96,boghosian1998quantum} that a quantum computer
can be programmed to {efficiently simulate the unitary time evolution of almost any physically realistic quantum system}.
The time evolution of a quantum system initialized in a given state $|\psi(0)\ra$ 
is governed by the Schr\"odinger equation
\begin{equation}
i \frac{d|\psi(t)\rangle}{dt}= \hat{H} |\psi(t)\rangle \quad,\quad t\ge 0 \quad,
\label{eq:SE}
\end{equation}
where $\hat{H}$ is the system's Hamiltonian. Since any fermionic or spin system
can be mapped to qubits, see \revision{Subsection}~\ref{sec:map2qubits}, below we assume that $\hat{H}$
describes a system of $n$ qubits. 
By integrating  Eq.~\eqref{eq:SE} for a time-independent Hamiltonian one obtains the time-evolved state
\begin{equation}
|\psi(t)\rangle=e^{-it \hat{H}} |\psi(0)\rangle \quad.
\label{eq:solSE}
\end{equation}
A quantum algorithm for {Hamiltonian simulation}
takes as input a description of $\hat{H}$, the evolution time $t$, and outputs
a quantum circuit $\hat{U}$ that approximates the time evolution 
operator $e^{-it \hat{H}}$ within a specified precision $\epsilon$, that is,
\begin{equation}
\label{Hsim}
\| \hat{U} - e^{-it \hat{H} }\|\le \epsilon \quad.
\end{equation}
More generally, the circuit $\hat{U}$ may use some ancillary qubits initialized in the $|0\ra$ state. 
The simulation cost is usually quantified by the runtime of the algorithm (the gate count of $\hat{U}$) and
the total number of qubits. 
Applying the circuit $\hat{U}$
to the initial state $|\psi(0)\rangle$ provides an $\epsilon$-approximation to the time-evolved
state $|\psi(t)\rangle$. 
The final state $\hat{U} |\psi(0)\ra$ can now be measured to access 
dynamical properties of the system such as time-dependent correlation functions.
The time evolution circuit $\hat{U}$ is usually invoked as a subroutine in a larger
enveloping algorithm.
For example, the quantum phase estimation method employs a controlled version of $\hat{U}$
to measure the phase accumulated during the time evolution, see \revision{Subsection}~\ref{sec:qpe}.
The enveloping algorithm is also responsible for preparing the initial state $|\psi(0)\rangle$.

While practical applications are concerned with specific Hamiltonian instances,
quantum simulation 
algorithms apply to general classes of Hamiltonians satisfying
mild technical conditions that enable a quantum algorithm to access the Hamiltonian
efficiently.
For example, a Hamiltonian can be specified  as a linear combination
of elementary
interaction terms denoted $\hat{V}_1,\ldots,\hat{V}_L$ such that
\begin{equation}
\label{Hlocal}
\hat{H} =\sum_{i=1}^{L} \alpha_i \hat{V}_i, \qquad \| \hat{V}_i\|\le 1.
\end{equation}
Here $\alpha_i$ are real coefficients and $\| \hat{V}_i \|$ is the operator norm (the maximum magnitude
eigenvalue).
In the case of \textit{local Hamiltonians}~\cite{Lloyd96,kitaev2002classical},
each term $\hat{V}_i$ acts non-trivially only a few qubits.
This includes an important special case of lattice Hamiltonians where
qubits are located at sites of a regular lattice and the interactions 
$\hat{V}_i$ couple small subsets of nearest-neighbor qubits. 
Molecular electronic Hamiltonians mapped to qubits assume the form Eq.~(\ref{Hlocal}), where
$\hat{V}_i$ are tensor products of Pauli operators which may have a super-constant weight.
This situation is captured by the 
\textit{Linear Combination of Unitaries} (LCU) model~\cite{Berry15a}.
It assumes that each term $\hat{V}_i$ is
a black-box unitary operator that can be implemented at a unit cost
by querying an oracle
(more precisely, one needs a ``select-$\hat{V}$" oracle 
implementing a
controlled version of $\hat{V}_1,\ldots,\hat{V}_m$).
The LCU model also assumes an oracle access to the coefficients $\alpha_i$,
see~\cite{Berry15a} for details.
Alternatively, a quantum algorithm can access the Hamiltonian through a subroutine
that computes its matrix elements. 
A Hamiltonian $\hat{H}$ is said to be {$d$-sparse}~\cite{Aharonov03} if the matrix
of $\hat{H}$ in the standard $n$-qubit basis
has at most $d$ non-zero entries in a single row or column.
The sparse Hamiltonian model assumes that 
positions and values of the nonzero entries 
can be accessed by querying suitable oracles~\cite{Berry15}.
Most  physically realistic quantum systems  can be mapped to
either local, LCU, or sparse qubit Hamiltonians such that the corresponding
oracles are realized by a short quantum circuit. As described below,
the runtime of quantum simulation algorithms 
is controlled by a dimensionless parameter $T$ proportional to the product of the evolution time
$t$ and a suitable norm of the Hamiltonian. One can view $T$ as an effective 
evolution time. A formal definition of $T$ for various Hamiltonian models is as follows.
\begin{equation}
\label{Teff}
T = \left\{ \begin{array}{rl}
t \max_i |\alpha_i| & \mbox{(Local)}\\
t\sum_i |\alpha_i| & \mbox{(LCU)} \\
td\|H\|_{max} & \mbox{(Sparse)} \\
\end{array} \right.
\end{equation}
Here $\|H\|_{max}$ denotes the maximum magnitude of a matrix element,
\revision{see \cite{childs2009limitations} for more extended discussions}.

It is strongly believed that the Hamiltonian simulation problem is hard for classical computers.
For example, Ref.~\cite{nagaj2008hamiltonian} showed that {\em any} problem solvable on a quantum computer
can be reduced to solving an instance of a suitably formalized
Hamiltonian simulation problem with a local Hamiltonian. 
Technically speaking, the problem is $\mathsf{BQP}$-complete~\cite{kitaev2002classical}.
All known classical methods capable of simulating general quantum systems of the above
form require resources (time and memory) exponential in $n$.
On the other hand, 
Feynman's original insight~\cite{feynman1982simulating} was that a quantum computer should be capable of
simulating many-body quantum dynamics  efficiently, such that the simulation runtime
grows only polynomially with the system size $n$ and the evolution time $T$. 
This intuition was confirmed by Lloyd who gave the first quantum algorithm for simulating local Hamiltonians \cite{Lloyd96}. The algorithm 
exploits the \textit{Trotter-Suzuki product formula}
\begin{equation}
e^{-i t ( \hat{H}_1+ \hat{H}_2+\ldots +  \hat{H}_L)} \approx 
\left(
e^{-i t \hat{H}_1}
e^{-i t \hat{H}_2}
\dots
e^{-i t \hat{H}_L}
\right)^{\frac{t}{\delta}}.
\end{equation}
By choosing a sufficiently small Trotter step $\delta$ one can approximate
the evolution operator $e^{-iHt}$  by a product of few-qubit operators 
describing evolution under individual  interaction terms.
Each few-qubit operator can be easily implemented by a short quantum circuit.
The runtime of Lloyd's algorithm scales as~\cite{Lloyd96,campbell2018random}
\begin{equation}
\label{Lloyd}
t_{run} = O(L^3 T^2 \epsilon^{-1}),
\end{equation}
where $T$ is the effective evolution time for local Hamiltonians, see Eq.~(\ref{Teff}).
 Importantly, the {runtime scales polynomially} with all relevant
parameters.

Lloyd's algorithm  was a  breakthrough result demonstrating that quantum computers can 
indeed provide an exponential speedup over the best known classical algorithms
for the task of simulating time evolution of quantum systems.
However, it was quickly realized that the runtime of  Lloyd's algorithm is unlikely to be optimal.
Indeed, since any physical system simulates its own dynamics in a real time,
one should expect that 
a universal quantum simulator can attain a runtime scaling only linearly with $t$.
Moreover, for any realistic Hamiltonian composed of short-range interactions on 
a regular lattice, one should expect that the simulation runtime is linear in the space-time
volume $nt$.  
Clearly, the scaling Eq.~(\ref{Lloyd}) falls far behind these expectations.

\subsection{Algorithmic tools}
\label{subs:algorithmic_tools}

The last decade has witnessed  several {improvements in the runtime scaling}
based on development of new algorithmic tools for Hamiltonian simulation.
Most notably, a breakthrough work by 
Berry et al.~\cite{Berry15,Berry15a,Berry13} 
achieved an {exponential speedup over Lloyd's algorithm
with respect to the precision $\epsilon$}.
A powerful algorithmic tool introduced in~\cite{Berry15}  is the so-called \textit{LCU lemma}~\cite{Berry15}.
It shows how to construct a quantum circuit that implements
an operator $\hat{U}^\prime =\sum_{i=1}^M \beta_i \hat{U}_i$, where $\beta_i$ are complex coefficients
and $\hat{U}_i$ are black-box unitary operators. Assuming that $U'$ is close to a
unitary operator, the lemma shows that $U'$ can be well approximated 
by a quantum circuit of size roughly $M\sum_i |\beta_i|$ using
roughly $\sum_i |\beta_i|$ queries to the oracle implementing $\hat{U}_1,\ldots,\hat{U}_M$
(and their inverses).
The simulation algorithm of 
Ref.~\cite{Berry15a} works by splitting the evolution into small intervals
of length $\tau$ and using 
the truncated Taylor series approximation $e^{-i\tau\hat{H}}\approx \sum_{m=0}^K (-i\tau\hat{H})^m/m!\equiv \hat{U}_\tau$.
Accordingly, $e^{-it\hat{H}}=(e^{-i\tau\hat{H}})^{t/\tau} \approx (\hat{U}_\tau)^{t/\tau}$.
Substituting the LCU decomposition of $H$ into the Taylor series
one obtains an LCU decomposition of $\hat{U}_\tau$. 
For a suitable choice of the truncation order $K$, the truncated series
$\hat{U}_\tau$ is close to a unitary operator. 
Thus $\hat{U}_\tau$ can be well approximated
by a quantum circuit using the LCU lemma. The runtime of this simulation
algorithm, measured by the number of queries to the Hamiltonian oracles,
scales as~\cite{Berry15a}
\begin{equation}
t_{run} = \frac{T\log{(T/\epsilon)}}{\log\log{(T/\epsilon)}}.
\end{equation}
Here $T$ is the effective evolution time for the LCU Hamiltonian model, see Eq.~(\ref{Teff}).
This constitutes a square-root  speedup with respect to $T$ and
an exponential speedup with respect to the precision compared with Lloyd's algorithm.

An important algorithmic tool proposed by Childs~\cite{Childs2010} is
converting a Hamiltonian into a \textit{quantum walk}. The latter is a unitary operator $W$
that resembles  the evolution operator $e^{-it \hat{H}}$ with a unit evolution time $t$.
For a suitable normalization of $\hat{H}$,
the quantum walk operator $\hat{W}$ has   eigenvalues $e^{\pm i \, \mathrm{arcsin}{(E_\alpha)}}$,
where $E_\alpha$ are eigenvalues of $\hat{H}$.
The corresponding eigenvectors of $\hat{W}$
are simply related to those of $\hat{H}$.
Unlike the true evolution operator, 
the quantum walk $\hat{W}$ can be easily  implemented using only a few
queries to the oracles describing the Hamiltonian $\hat{H}$,
e.g. using the LCU or the sparse models.
To correct the discrepancy between $\hat{W}$ and the true evolution operator, 
Low and Chuang~\cite{low2017optimal} proposed the \textit{Quantum Signal Processing} (QSP) method. 
One can view QSP as a compiling algorithm that takes as input a black-box unitary 
operator $W$, a function $f\, : \, \CC\to \CC$,  and outputs a quantum circuit that realizes $f(\hat{W})$.
Here it is understood that $f(\hat{W})$ has the same eigenvectors as $\hat{W}$ while each eigenvalue
$z$ is mapped to $f(z)$.
The  circuit realizing $f(\hat{W})$ is expressed using controlled-$\hat{W}$ gates and single-qubit gates on the control qubit.
Remarkably,  it can be shown that the  Low and Chuang algorithm is {optimal}  for
the sparse Hamiltonian model~\cite{low2017optimal}.  Its runtime, measured by the number of queries
to the Hamiltonian oracles, scales as 
\begin{equation}
t_{run} = T +\frac{\log(1/\epsilon)}{\log\log(1/\epsilon)}
\label{QSP}
\end{equation}
where  $T$ is the effective evolution time for the sparse model, see Eq.~(\ref{Teff}).
This scaling is optimal in the sense that it matches  previously known lower bounds~\cite{Berry13,Berry15}.
We note that 
simulation methods based on the sparse and LCU Hamiltonian models have been recently
unified using a powerful framework known as \textit{qubitization}~\cite{low2019hamiltonian-2}. It provides
a general recipe for converting a Hamiltonian into a quantum walk
using yet another oracular representation
of a Hamiltonian known as a block encoding~\cite{low2019hamiltonian-2}.

Algorithms based on the quantum walk (such as the QSP) 
or truncated Taylor series  
may not be the best choice for 
near-term applications since they require {many ancillary qubits}.

In contrast, the original Lloyd algorithm~\cite{Lloyd96} and its generalizations based on
\textit{higher order product formulas}~\cite{BACS07} require only as many qubits as needed
to express the Hamiltonian.
In addition, such algorithms are {well-suited for simulating lattice Hamiltonians} where 
qubits are located at sites of a regular $D$-dimensional grid and
each elementary interaction $V_i$ couples a few qubits located nearby.
Lattice Hamiltonians contain $L=O(n)$ elementary interactions.
Each few-qubit operator that appears in a product formula
approximating $e^{-it\hat{H}}$ can be expressed using a few gates that couple nearest-neighbor qubits.
The corresponding quantum circuit can be easily implemented on a device whose
qubit connectivity graph is a $D$-dimensional grid.

Quite recently, Childs and Su~\cite{childs2019nearly} revisited simulation algorithms based
on product formulas and demonstrated 
that their {performance is better than 
what one could expect from naive error bounds}.
More precisely, an order-$p$ product formula approximates the evolution
operator $e^{-iHt}$ 
under a Hamiltonian $\hat{H}=\hat{A}+\hat{B}$ with
a simpler operator that involves time evolutions under Hamiltonians $A$ and $B$ such that
the approximation error scales as $t^{p+1}$ in the limit $t\to 0$. 
Childs and Su~\cite{childs2019nearly}  showed 
that the gate complexity of simulating a lattice Hamiltonian using order-$p$ product formulas scales as 
$(nT)^{1+1/p}\epsilon^{-1/p}$ which shaves off a factor of $n$ from the best previously known bound.
Here $T$ is the effective evolution time for the local Hamiltonian model.

\begin{figure*}
    \centering
    \includegraphics[width=0.6\textwidth]{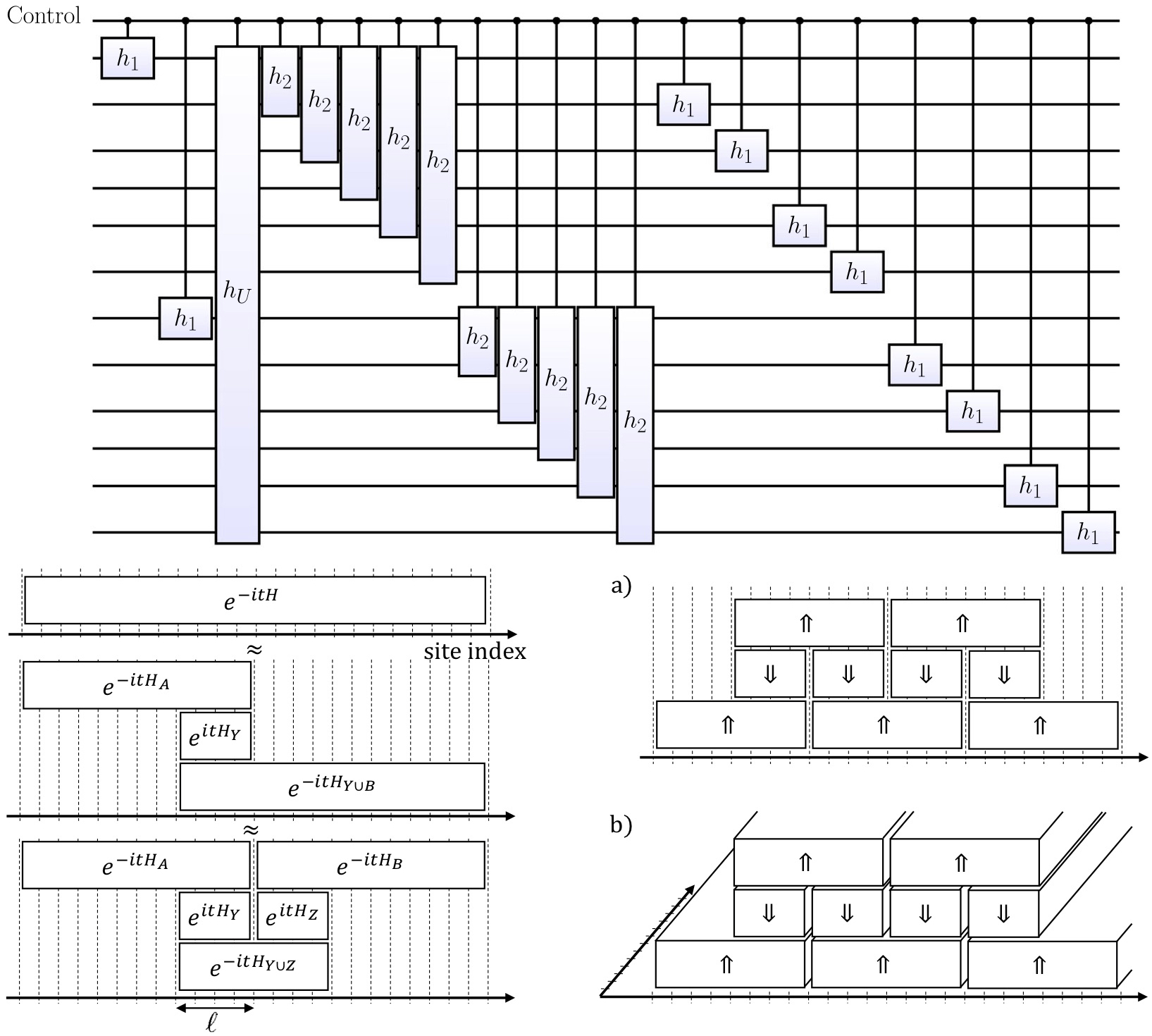}
    \caption{Trotter-like time evolution (top) and divide-and-conquer strategy for lattice systems (bottom). From Ref.~\cite{bauer2016hybrid} and Ref.~\cite{haah2018quantum} respectively.}
    \label{fig:diairei}
\end{figure*}

\revision{Haah et al.~\cite{haah2018quantum}.
recently introduced} a new class of {product formulas 
based on the \textit{divide-and-conquer} strategy}
and showed that lattice Hamiltonians can be simulated with gate complexity
$\tilde{O}(nT)$, where $\tilde{O}$ hides logarithmic corrections. 
This  result confirms the physical intuition that the cost of simulating lattice Hamiltonians
scales linearly with the space-time volume.
The algorithm of Ref.~\cite{haah2018quantum}, sketched in \revision{Figure}~\ref{fig:diairei}, approximates
the full evolution operator by dividing the lattice into small (overlapping) subsystems
comprising $O(\log{n})$ qubits each and simulating time evolution of  the individual subsystems.
The errors introduced by truncating the Hamiltonian
near the boundaries are canceled by alternating between forward and backward
time evolutions. The error analysis is based on a skillful application of the \textit{Lieb-Robinson 
bound}~\cite{lieb1972finite,hastings2006spectral}
that controls how fast information can propagate across the system during the time evolution.

Finally, we note that while product formulas achieve a better scaling with system size than LCU or QSP methods by exploiting commutativity of Hamiltonian terms, they suffer from worse scaling with the simulation time and the error tolerance. 
A recent approach of \textit{multiproduct formulas}~\cite{low2019well} combines the best features of both. From Trotter methods, it inherits the simplicity, low-space requirements of its circuits, and a good scaling with system size. From LCU, it inherits the optimal scaling with time and error, up to logarithmic factors. Essentially, the work~\cite{low2019well} shows how a certain type of high-order product formula can be implemented with a polynomial gate cost in the order (scaling as $p^2$). In contrast, the standard Trotter-Suzuki formulas scale exponentially with the order (scaling as $5^p$).

\subsection{Open problems}

The existing simulation methods such as QSP are {optimal in terms of the
query complexity}. However their {runtime may or may not be optimal} if one accounts for the cost 
of implementing the Hamiltonian oracles by a quantum circuit. 
Indeed, in the case of lattice Hamiltonians, \textit{system-specific simulation strategies}
 that are not based on oracular models are known to achieve better runtime~\cite{haah2018quantum}. 
 Improved system-specific simulation methods are also available for quantum chemistry systems.
For example Ref. \cite{babbush2018low} 
achieved a quadratic reduction in the number of interaction terms
present in molecular Hamiltonians by treating the kinetic energy and the potential/Coulomb
energy operators using two different sets of basis functions --  the plane wave basis and its dual. 
The two bases are related by the fermionic version of the Fourier transform
which admits a simple quantum circuit~\cite{babbush2018low}.
This simplification was shown to reduce the depth of simulation circuits
(parallel runtime)
based on the Trotter-Suzuki and LCU decompositions~\cite{babbush2018low}.
One may anticipate that further improvements can be made by
{exploiting system-specific information}.
It is an interesting open question whether molecular Hamiltonians
can be simulated in depth scaling  poly-logarithmically with the size
of the electronic basis.
More recent advanced sparse Hamiltonian simulation algorithms can exploit prior knowledge of other Hamiltonian norms to get better performance. For instance, the algorithm of Ref.~\cite{low2017hamiltonian}
has an effective time $T = \sqrt{d\|H\|_{\text{max}}\|H\|_1}$, which is a tighter bound than the 
one displayed in Eq.~(\ref{Teff}). This can be further improved to  
$T=t\sqrt{d}\|H\|_{1\rightarrow 2}$ which is a tighter bound than both, see~\cite{low2019hamiltonian} 
for details.
However, these scaling improvements come with a larger constant factor that 
has not been thoroughly characterized.

In many situations the Hamiltonian to be simulated can be written as $H=H_0+V$,
where the norm of $H_0$ is much larger than the norm of $V$, while
the time evolution generated by $H_0$ can be 
``fast-forwarded'' by calculating the evolution operator analytically.
For example, $H_0$ could represent the kinetic energy in the momentum basis
or, alternatively, potential energy in the position basis.
\textit{Hamiltonian simulation in the interaction picture}~\cite{low2018hamiltonian} is an algorithmic 
tool proposed to take advantage of such situations.
It allows one to pay a logarithmic cost (instead of the usual linear cost) with respect to the norm 
of $H_0$. This tool has been used to simulate chemistry in the plane wave basis with $O(N^2)$ 
gates in second quantization, where $N$ is the number of plane waves~\cite{low2018hamiltonian}.
In the first-quantization, the gate cost scales as $O(N^{1/3} K^{8/3})$,
where $K$ is the number of electrons~\cite{babbush2018quantum}.

Likewise, better bounds on the runtime may be obtained
by {exploiting the structure of the initial state} $\psi(0)$. In many applications, $\psi(0)$
is the Hartree-Fock approximation to the true ground state. As such, the energy of $\psi(0)$
tends to be small compared with the full energy scale of $\hat{H}$ and one may expect that the
time evolution is confined to the low-energy subspace of $\hat{H}$.
How to develop state-specific simulation methods and improved bounds on
the runtime is an intriguing  open problem posed in~\cite{HaahQIP2019}.

An interesting alternative to quantum simulations proposed by Poulin et al.~\cite{poulin2018quantum}
is {using the quantum walk operator} $\hat{W}$ described above directly {in the quantum phase estimation}
method to estimate eigenvalues of $\hat{H}$. This circumvents errors introduced by 
the Trotter or LCU decomposition reducing the total gate count. 
It remains to be seen whether other physically relevant quantities such as time
dependent correlation functions can be extracted from the quantum walk operator
sidestepping the unitary evolution which is relatively expensive 
compared with the quantum walk.

While the asymptotic runtime scaling is of great theoretical interest,
practical applications are mostly concerned with specific  Hamiltonian instances
and constant approximation error (e.g. four digits of precision).
To assess the practicality of quantum algorithms
for specific problem instances
and compare their runtime with that of state-of-the-art classical algorithms,
one has to examine 
 \textit{compiling methods} that transform
a high-level description of a quantum algorithm into a quantum circuit
that can be run on  particular hardware. 
This motivates development of compilers tailored to quantum
simulation circuits. For example, 
\revision{Whitfield et al.~\cite{whitfield2010}} and Hastings et al.~\cite{Hastings15} examined
Trotter-Suzuki type simulation of chemical Hamiltonians
mapped to qubits using the Jordan-Wigner transformation.
An improved compiling method was proposed 
reducing the runtime  
by a factor $O(N)$, where $N$ is the number of orbitals.
Childs et al.~\cite{childs2018faster} and Campbell~\cite{campbell2018random}
recently  proposed a randomized compiler tailored to Trotter-Suzuki simulations.
It improves the asymptotic runtime scaling Eq.~(\ref{Lloyd}) to
$L^{5/2}T^{3/2}\epsilon^{-1/2}$ and achieves 
almost $10^5$ speedup for simulation of small molecule quantum chemistry Hamiltonians~\cite{childs2018faster,campbell2018random}.
Different types of  compilers may be needed for NISQ devices
and fault-tolerant quantum machines~\cite{nam2019low}.
Indeed, in the context of NISQ devices,
arbitrary single-qubit gates are cheap and the simulation cost is dominated 
by the number of two-qubit gates. In contrast, the cost of implementing an
error-corrected quantum circuit is usually dominated by the number of non-Clifford
gates such as the T-gate or the CCZ-gate. Since large-scale quantum simulations
are expected to require error correction, developing compiling algorithms minimizing
the T-gate count is vital. For example, Low et al.~\cite{low2018trading} recently
achieved a square-root reduction in the T-count for a state preparation subroutine
employed by the LCU simulation method.
Compiling algorithms that minimize the number of generic single-qubit
rotations (which are expensive to implement fault-tolerantly)
were investigated by Poulin et al.~\cite{poulin2018quantum}.
This work considered implementation of the quantum walk operator $W$
associated with a lattice Hamiltonian that contains only a few distinct parameters
(e.g. translation invariant models).
\revision{Other important works that focused on the problem of optimizing circuits for quantum chemistry simulations include \cite{jones2012faster,yung2014transistor}.}

Estimating resources required to solve practically important problems
is linked with the study of \textit{space-time tradeoffs} in quantum simulation~\cite{bennett1989time,parent2015reversible,meuli2019reversible}.
For example, in certain situations the circuit depth (parallel runtime) can be reduced
at the cost of introducing ancillary qubits~\cite{barenco1995elementary} and/or using intermediate measurements
and feedback~\cite{gottesman1999demonstrating}.
Of particular interest are methods
for reducing the size of quantum simulation circuits
using ``dirty'' ancilla --- qubits 
whose initial state is unknown and which must be restored to their original form upon the completion of the 
algorithm~\cite{barenco1995elementary,haner2016factoring,low2018trading}, see also~\cite{buhrman2014computing}.
For example, if a Hamiltonian simulation circuit is invoked as a subroutine from 
a larger enveloping algorithm, the role of dirty ancillas may be played by data qubits
borrowed from  different parts of the algorithm.

The Hamiltonian simulation problem has a purely classical version.
It  deals with the
ODE $\dot{\mathbf{p}}=-\partial H/\partial {\mathbf{q}}$, $\dot{{\mathbf{q}}}=\partial H/\partial {\mathbf{p}}$,
where $\mathbf{p},\mathbf{q}$ are canonical coordinates of a classical Hamiltonian system.
Numerical algorithms for integrating classical Hamiltonian dynamics known as \textit{symplectic integrators}
have a long history~\cite{channell1990symplectic,ruth1983canonical} 
and are widely applied in simulations of molecular dynamics~\cite{gray1994symplectic}.
One may ask whether new advances in quantum or classical Hamiltonian simulation
algorithms can be made by an exchange of ideas between these
two fields.

Finally, the problem of \textit{simulating Hamiltonians in the presence of noise}, without active error correction in the NISQ era,
is an open problem. The primary question in this context is whether noise tolerant methods, analogous to variational algorithms in ground-state
energy optimization, can also be developed for Hamiltonian simulation.

\subsection{Finite-temperature algorithms}
\label{sec:quantum_thermal_states}

How a quantum computer can be used to simulate experiments 
on quantum systems in {thermal equilibrium} is an important 
problem in the field of quantum simulation. Early algorithms for the simulation of Gibbs states
\cite{Terhal_PRA_2000,Poulin2009,Eisert2012} were
based on the idea of coupling the system of interest 
to a set of ancillary qubits and letting the system and 
bath together evolve under a joint Hamiltonian, thus mimicking
the physical process of thermalization. 
The main disadvantages of thermalization-based methods 
are the presence of additional ancillary qubits defining
the bath states, and the need to perform time evolution
under $\hat{H}$ for a thermalization time $t$ that could 
be very long.

More recent proposals have focused on 
ways to generate finite-temperature observables without 
long system-bath thermalization times and/or large ancillae bath representations.
For example, Ref.~\cite{chowdhury2016quantum} showed how to realize the imaginary time evolution operator $e^{-\beta \hat{H}/2}$
using  Hamiltonian simulation techniques. 
Applying  a suitable version of the Hubbard-Stratonovich transformation
the authors of Ref.~\cite{chowdhury2016quantum} obtained a representation 
$e^{-\beta \hat{H}/2}=\sum_\alpha c_\alpha U_\alpha$,
where $c_\alpha$ are real coefficients and $U_\alpha$ 
are unitary operators describing the time evolution under
a Hamiltonian $\hat{H}^{1/2}$. 
Although the square-root $\hat{H}^{1/2}$ is generally not easily available, Ref.~\cite{chowdhury2016quantum}
showed how to realize it for Hamiltonians composed of few-qubit positive semidefinite
terms using an ancillary system. Applying state-of-the-art methods to simulate
time evolution under $\hat{H}^{1/2}$ and the LCU Lemma (see \revision{Subsection}~\ref{subs:algorithmic_tools}) to realize 
the desired linear combination of unitaries, Ref.~\cite{chowdhury2016quantum} obtained a quantum 
algorithm for preparing the thermal Gibbs state with gate complexity
$\beta^{1/2} 2^{n/2} {\cal Z}^{-1/2}$, where $\cal Z$ is the quantum partition function.
Here we ignored a prefactor scaling poly-logarithmically with $\beta$ and the inverse
error tolerance. A closely related but slightly less efficient algorithm was discussed in Ref.~\cite{poulin2009sampling}.

Several more heuristic quantum algorithms for finite-temperature simulations have been 
proposed recently. However, most of these algorithms are challenging to analyze mathematically and generally do not have performance guarantees.
For example, \textit{variational ansatz states for the Gibbs state} that can be prepared with simple circuits
have been proposed to bypass possibly long thermalization times. One example is the \textit{product spectrum ansatz} \cite{Martyn_arxiv_2018} (PSA),
where a shallow unitary circuit applied to a product thermal state is chosen to minimize the free energy of the system.

\begin{figure}
    \centering
    \includegraphics[width=0.75\columnwidth]{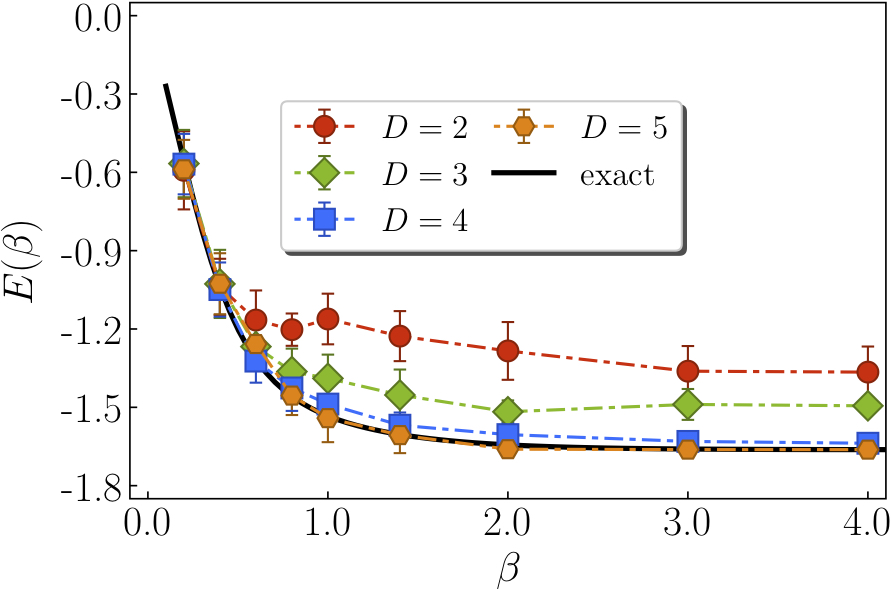}
    \caption{Application of the quantum METTS algorithm to the 6-site 1D Heisenberg model. Energy is shown as a function of inverse temperature $\beta$, using unitaries acting on $D=2 \dots 5$ spins. From \cite{Motta2019}.}
    \label{fig:qmetts}
\end{figure}

A different avenue is to sample from the Gibbs state rather than generate it explicitly on the quantum computer.
For example, \textit{quantum Metropolis sampling}~\cite{temme2011quantum,yung2012quantum}  samples from the Gibbs state in an
analog of classical Metropolis sampling,  using phase estimation on a random unitary applied
to the physical qubits to ``propose'' moves, and an iterative amplification procedure to  implement the ``rejection''.
Much like the classical Metropolis algorithm, the fixed point of this procedure samples the Gibbs state.
Alternatively, the \textit{quantum minimally entangled typical thermal state (METTS) algorithm}~\cite{Motta2019} samples from the Gibbs state using imaginary time evolution applied to pure states, implemented via the quantum imaginary time evolution algorithm. One strength of quantum METTS is that it uses only the physical qubits of the system and potentially shallow circuits, thus making it feasible even in the NISQ era, where it has been demonstrated
on quantum hardware for small spin systems.
An application to the Heisenberg model is shown in \revision{Figure}~\ref{fig:qmetts}.

Another alternative is to work within the microcanonical ensemble. This is the basis
of the \textit{minimal effective Gibbs ansatz (MEGA)}~\cite{Cohn_arxiv_2018}, which attempts to generate pure states within the energy window corresponding
to a microcanonical ensemble (for example, using phase estimation). The basic challenge is to ensure
that the energy window is chosen according to the desired temperature. The MEGA method
estimates the temperature from asymptotic properties of the ratio
of the greater and lesser Green's functions, which in principle can be measured on the quantum device
using the techniques in \revision{Section}~\ref{sec:readout}.

While there are many different proposed techniques for estimating observables of thermal states which all
appear quite plausible on theoretical grounds, little is known about their heuristic performance, and
almost none have been tested on real devices. In this sense, thermal state algorithms lag greatly behind
those for ground-states for problems of interest in chemistry and materials science.
To identify the best way forward, heuristic benchmarking for systems of relevance in physics and chemistry will be of major importance.
The problem of benchmarking is further discussed in \revision{Section}~\ref{sec:benchmarks}. 

\subsection{Resource estimates}

\revision{An important task in the practical demonstration and application of quantum 
algorithms, especially on contemporary hardware, is to determine and optimize the amount of quantum 
resources required. Several groups have provided careful quantitative estimates of these resources, expressed in 
terms of parameters such as the number of qubits, the number of single-qubit 
and $\mathsf{CNOT}$ gates, the circuit depth and the number of measurements
required by an algorithm. These are listed in a number of recent works, and for a
variety of quantum algorithms \cite{aspuru2005simulated,reiher2017elucidating,childs2018toward,Motta2019,motta2020r12f12,babbush2018low,wecker2015progress,lemieux2020resource,kivlichan2019improved}.
These estimates not only provide a snapshot of current trends, but also
indicate that quantum simulations can be very expensive, with
estimated circuit complexities for simulating interesting problems still many orders of magnitude
greater than what can be achieved with contemporary hardware. While the resource estimates
can be expected to come down as better optimized algorithms are developed, and hardware
resources are increasing with time, the need to share the computational cost of a simulation between
scarce quantum resources and abundant classical resources, especially by leveraging the structure of the
problem at hand, is the motivation for developing hybrid quantum-classical methods. Some of these have
already been discussed above, and further examples are given in the next section.}

\subsection{Hybrid quantum-classical methods}

\label{sec:quantum_classical}

\subsubsection{Quantum embedding methods}

Embedding algorithms use a divide-and-conquer strategy to {break a large quantum simulation into smaller pieces} that are more amenable to simulation. The properties of the original model and the reduced models are related to each other in a self-consistent fashion.
These methods are popular both in {condensed matter physics} to study correlated electronic materials,
where they reduce the problem of solving a bulk fermionic lattice model to that of studying a simpler Anderson-like impurity model, 
as well as in {molecular applications}, to reduce the computational scaling of methods in the
simulation of complicated molecules.

In classical simulations, there are {many flavors} of quantum embedding. These can be grouped roughly by
the choice of variable used to communicate between regions; \textit{dynamical mean-field
theory} (DMFT) works with the Green's function and self-energy~\cite{kotliar2006electronic, GeorgesDMFT,sun2016quantum}; \textit{density matrix embedding theory} with the one-body density matrix~\cite{knizia2012density,sun2016quantum}; \textit{density functional embedding} via the electron density~\cite{wesolowski1993frozen,sun2016quantum}, and other methods, such as QM/MM, ONIOM, and fragment MO methods communicate via the electrostatic potential~\cite{senn2009qm}. 

\begin{figure}
    \centering
    \includegraphics[width=0.75\columnwidth]{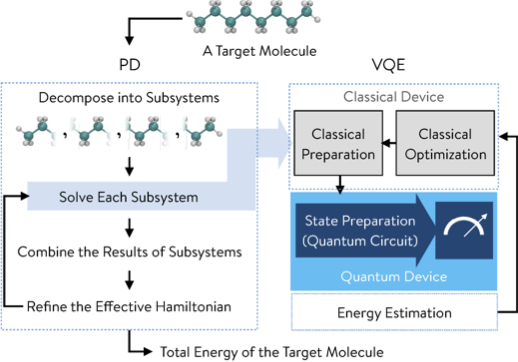}
    \caption{Illustration of hybrid quantum-classical embedding for a quantum chemistry system. Fragmentation of the original problem is performed on a classical computing device, and the more complex task of simulating each subproblem is handled by the quantum algorithm. From Ref.~\cite{yamazaki2018towards}.}
    \label{fig:embedding}
\end{figure}

There has been {growing interest in quantum embedding} methods in the quantum information community,
with the quantum computer playing the role of the quantum mechanical solver for the fragment/impurity problem (see \revision{Figure}~\ref{fig:embedding}).
Ref.~\cite{bauer2016hybrid} suggested that a small quantum computer with a few hundred qubits
could potentially speed up material simulations based on the DMFT method, and proposed a quantum algorithm for computing 
the Green's function of a quantum impurity model. Kreula et al. \cite{kreula2016few} subsequently proposed a proof-of-principle demonstration of this  algorithm. Similarly, Rubin~\cite{rubin2016hybrid} and Yamazaki et al~\cite{yamazaki2018towards} have explored the potential of DMET in conjunction
with a quantum computer for both condensed-phase lattice models as well as for large molecular calculations. 
Ground states of quantum impurity models and their structural properties in the context
of quantum algorithms have been analyzed in Ref.~\cite{bravyi2017complexity}. 

\subsubsection{Other hybrid quantum-classical algorithms}

Beyond standard quantum embedding, there are many other \revision{possibilities} for hybrid quantum-classical algorithms.
Variational quantum eigensolvers have previously
been discussed for eigenstate problems, see \revision{Subsection}~\ref{sec:vqe}. One can also classically postprocess a quantum simulation of
an eigenstate to improve it, in an analog of post-Hartree-Fock and post-complete-active-space methods, as has been explored
in \cite{takeshita2019increasing,motta2020r12f12}.
In quantum molecular dynamics, it is natural to use
a quantum computer to propagate the wavefunction or density matrix subject
to motion of the nuclei, as is done today in classical Born-Oppenheimer dynamics. Similarly, the use of quantum optimizers and quantum annealers, for example to assist classical conformational search~\cite{marchand2018variable}, can also be viewed as types of quantum-classical hybrids.

\subsection{Open questions}

It is clear that one will {rely on hybrid quantum-classical algorithms for many years to come}, and there remain many open questions.
One is how to {best adapt quantum algorithms} within existing quantum-classical frameworks. For example,  Green's function embedding
methods are generally formulated in terms of actions rather than Hamiltonians; unfolding into a bath representation, consuming
additional qubits, is currently required. More compact representations of the retarded interactions suited for quantum simulation should be explored
(see for example, Ref.~\cite{arguello2018analog} for a related proposal to generate effective long-range interactions). Similarly one should explore
the best way to evaluate Green's functions or density matrices, minimizing the coherence time and number of measurements.
It is also possible that new kinds of embedding frameworks should be considered.
For example, \textit{quantum-quantum embedding} algorithms within the circuit model
of quantum computation have been proposed to simulate
large-scale ``clustered'' quantum circuits on a small quantum computer~\cite{bravyi2016trading,peng2019simulating}.
These are circuits that can be divided into small clusters
such that there are only a few entangling gates connecting different clusters.
Another promising class of embedding algorithms known
as \textit{holographic quantum simulators} was also recently proposed in~\cite{kim2017holographic,kim2017noise}.
Such algorithms enable the simulation of 2D lattice models on a 1D quantum computer by converting one spatial dimension into time.
Whether it is useful to incorporate a classical component into such quantum-quantum frameworks clearly needs to be explored. 

Another question is how best to implement the \textit{feedback between the quantum and classical parts} of the algorithm. For example,
as already discussed in the variational quantum eigensolver, such optimizations require the evaluation of approximate gradients on the quantum
device, and  noisy  optimization with limited gradient information on the classical device. Improving both aspects is clearly needed.

\section{Benchmark systems}
\label{sec:4}

\label{sec:benchmarks}

Researchers working on quantum simulation algorithms would greatly benefit
from having access to {well-defined benchmarks}. Such benchmarks help the community by
defining common conventions (e.g. choosing specific bases) and curating the best results.

There are two types of benchmarks to develop.
\begin{itemize}
\item the first is a benchmark that allows {quantum algorithms}
(possibly on different hardware) to be compared {against each other}. For example, such
problems could include a test suite of molecular Hamiltonian simulation problems
for some specific choice of electron basis functions, fermion-to-qubit mapping,
evolution time, and the desired approximation error.
\item The second is curated {data from the best classical methods}
for specific problems and well-defined Hamiltonians. Wherever possible, the data should not only include
ground-state energies, but also excited-states and other observables, and if exact results are not available,
an estimate of the precision should be given.
\end{itemize}

In the near-term era, suitable candidates for benchmarking may include molecular or material science Hamiltonians that can be expressed with about 50 or
fewer qubits. Some promising  candidates discussed in the literature include lattice spin Hamiltonians~\cite{childs2018toward},
and models of correlated electrons such as the 2D Fermi-Hubbard model~\cite{wecker2015progress},
the uniform electron gas (jellium)~\cite{babbush2018low}, and the Haldane pseudo-potential Hamiltonian that models FQHE systems~\cite{chen2015algebraic,dunne1993slater,di2017unified}. \revision{In molecular simulations, small transition metal compounds, such as the chromium dimer,
  have been the subject of extensive benchmarks in small and manageable basis sets and thus
  are a potential benchmark also for quantum algorithms \cite{williams2020direct,olivares2015ab}.}
While some aspects of these models are easy to solve classically for systems of moderate size,
others remain difficult, providing room for quantum advantage.

Naturally, there is a wide range of molecular or materials problems that could be chosen, a small number of which are highlighted in
\revision{Section}~\ref{sec:simulations}. One relevant factor to check when constructing a Hamiltonian benchmark problem is to verify that
it is indeed difficult to simulate classically~\cite{li2019electronic2}.
Ideally, there should be a way to maximally tune the ``complexity'' for classical simulation, which then defines a natural setting for demonstrating quantum supremacy.
Commonly, a way to tune the Hamiltonian (in model Hamiltonians) is to change the parameters of the Hamiltonian directly. In more realistic settings, one may  change the size or geometry of the system or the chemical identity of the atoms. 

\section{Reading out results}
\label{sec:5}

\label{sec:readout}

\subsection{Equal-time measurements}

It is of course an essential part of any quantum computation to perform a {measurement on the final state} and thus {read out the result} of the computation. Conventionally, this is achieved through projective measurements of individual qubits in the computational basis. More complicated operators can be measured through standard techniques, for example by first applying a unitary rotation to the state that maps the operator onto Pauli-$Z$, or by using an ancillary qubit.

For very {complicated operators}, however, this can become quite {resource-intensive}. Consider for example measuring the expectation value of the Hamiltonian, which is required e.g. in the VQE (see \revision{Subsection}~\ref{sec:vqe}). This can be done by writing the Hamiltonian as a sum of products of Pauli operators and measuring each one individually. Each measurement must be repeated a sufficient number of times to collect accurate statistics. Since for many applications, the number of operators grows quite quickly with the number of qubits (for example as $N^4$ in typical quantum chemistry applications) and the state may have to be prepared anew after a projective measurement, it is important to {organize the terms} in such a way that the number of operations to achieve a desired accuracy is minimized. Some work in this direction appears in Sec. V of Ref.~\cite{kandala2017hardware}, as well as more recently in Refs.~\cite{jena2019pauli,verteletskyi2019measurement}. \revision{Other approaches have been very recently proposed, based on grouping terms into parts whose eigenstates have a single-qubit product structure and devising multi-qubit unitary transformations for the Hamiltonian or its parts to produce less entangled operators \cite{izmaylov2019revising,zhao2020measurement}, as well as on the use of neural network techniques to increase the precision of the output results \cite{torlai2020precise}. While these results are encouraging, the need to converge a large number of measurements
to high precision remains a practical problem in many implementations, particularly in many hybrid quantum-classical algorithms.}

\subsection{Dynamical properties and Green's functions}

Much of the experimentally relevant information about a system, for example as obtained in scattering experiments such as optical or X-ray spectroscopy or neutron scattering, is encoded in {dynamical properties}. Access to these properties allows for a more direct comparison between theoretical predictions and experiments, thus allowing to infer microscopic information that is difficult to extract solely from the experiment.

\begin{figure}
    \centering
    \includegraphics[width=0.75\columnwidth]{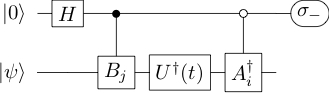}
    \caption{Quantum circuit to measure $\langle \Psi | \hat{A}^\dag_i(t) \hat{B}_j | \Psi \rangle$. Here $\hat{A}_i$, $\hat{B}_j$ are unitaries, $\hat{A}^\dag_i(t) = \hat{U}(t) \hat{A}^\dag_i \hat{U}^\dag(t)$ is the time evolution of $\hat{A}^\dag_i$ under $\hat{U}(t) = e^{-i t \hat{H}}$, and $2 \sigma_- = 2 | 0 \rangle \langle 1 | = X+iY$.
    Adapted from Ref.~\cite{somma2003quantum}.}
    \label{fig:green_circuit}
\end{figure}

A convenient way to capture this information is via the single- or few-particle \textit{Green's functions}, which can be simply
related to time-correlation functions of observables, such as the dynamic structure factors and dipole-dipole correlation functions.
For example, the particle and hole components of the single-particle Green's function in real time ($t \geq 0$) are given by
\begin{equation}
\begin{split}
G^p_{\alpha \beta}(t) &= \langle \psi | \hat{c}_\alpha(t) \hat{c}_\beta^\dagger(0) |\psi\rangle \;, \\
G^h_{\alpha \beta}(t) &= \langle \psi | \hat{c}_\alpha(t)^\dagger \hat{c}_\beta(0) |\psi\rangle \;, \\
\end{split}
\end{equation}
where $\alpha,\beta$ can be spin, orbital or site indices, $|\psi\rangle$ is the quantum state of interest (for example the ground state), and $\hat{c}_\alpha^{(\dagger)}(t) = e^{it \hat{H}} \hat{c}_\alpha^{(\dagger)} e^{-it \hat{H}}$. These can be measured by decomposing the fermion creation and annihilation operators into unitary combinations and applying standard techniques, see e.g. Ref.~\cite{somma2003quantum,bauer2016hybrid} (a quantum circuit for measuring products of unitaries, one of which time-evolved, is shown in \revision{Figure}~\ref{fig:green_circuit}. 

However, this procedure is {expensive} as it requires separate calculations for each time $t$, the particle/hole, real/imaginary and potentially orbital/spatial components of the Green's function, and finally a potentially large number of repetitions for each to achieve some desired target accuracy. While some initial work has shown how to improve this slightly by avoiding having to re-prepare the initial state every time (which could be prohibitive), further improvements and developments of
alternatives, would be very valuable.

It should be noted that the behavior of the Green's function is very constrained both at short times (due to sum rules) and at long times (where the decay is governed by the longest time-scale of the system). Therefore, of primary interest is the {regime of intermediate times} where classical methods are most difficult to apply. However, in these other regimes, the additional structure could potentially be employed to reduce the computational effort also on a quantum computer.

\section{Conclusions}
\label{sec:6}

In this review we surveyed some of the quantum problems
of chemistry and materials science that pose a challenge for classical computation, 
and the wide range of quantum algorithms that can potentially target such
problems. 
The field of quantum algorithms for physical simulation is growing at an exponential pace, thus this
review  only provides a snapshot of progress. However, there are many shared ideas and components
and we have tried to bring out these common features, as well
as highlight the many open problems that remain to be explored.

Although very few experiments have yet been carried out in actual quantum hardware, it will soon become possible to
test many proposals in real quantum computational platforms. While it is not at all certain then which techniques will perform best in practice,
what is clear is that such experiments will greatly enhance our understanding of the heuristics of quantum algorithms.
Indeed, the widespread availability of classical
computers led to a new wave of algorithmic innovation, and the establishment of computational chemistry, materials science, and physics
as scientific fields. Should quantum hardware
continue to develop at the current rate, we can thus look forward also to a new wave of developments in quantum algorithms,
and the emergence of new computational disciplines dedicated to the study of chemistry, materials science, and physics on quantum computers.

\section*{Acknowledgement}
This review was adapted and condensed from a report produced for a US National Science Foundation workshop
held between January 22-24 2019 in Alexandria VA. Funding for the workshop and for the production of the initial report was provided by the US National Science Foundation
via award no. CHE 1909531.

Additional support for GKC for the production of this review from the initial report
was provided by the US National Science Foundation via
award no. 1839204.

The workshop participants included:
 Garnet Kin-Lic Chan,
 Sergey Bravyi,
 Bela Bauer,
 Victor Batista,
 Tim Berkelbach,
  Tucker Carrington,
 Gavin Crooks,
 Francesco Evangelista,
  Joe Subotnik,
   Haobin Wang,
 Bryan Clark,
 Jim Freericks,
  Emanuel Gull,
 Barbara Jones,
  Austin Minnich,
 Steve White,
  Andrew Childs,
    Sophia Economou,
 Sabre Kais,
 Guang Hao Low,
 Antonio Mezzacapo,
 Daniel Crawford,
 Edgar Solomonik,
 Takeshi Yamazaki,
Christopher Chang, Alexandra Courtis, Sarom Leang, Mekena Metcalf, Anurag Mishra, Mario Motta, Petr Plechac, Ushnish Ray, Julia Rice, Yuan Su, Chong Sun, Miroslav Urbanek, Prakash Verma, Erika Ye.
The workshop website is: {https://sites.google.com/view/nsfworkshopquantumleap/}.

\end{document}